\documentstyle[eqsecnum,aps,prb,multicol,epsf]{revtex}
\begin{document}

\draft
%\tighten
\title{\hfill {\small ITP Preprint Number NSF-ITP-97-063}  \\
\vspace{10pt}
Delocalization Transition via Supersymmetry in
One-Dimension
}
\author{Leon Balents and  Matthew P. A. Fisher}
\address{Institute for Theoretical Physics, University of California, 
Santa Barbara, CA 93106--4030}

\date{\today}
\maketitle

%\widetext

\begin{abstract}
  We use supersymmetric (SUSY) methods to study the delocalization
  transition at zero energy in a one-dimensional tight-binding model
  of spinless fermions with particle-hole symmetric disorder.  Like
  the McCoy-Wu random transverse-field Ising model to which it is
  related, the fermionic problem displays two different correlation
  lengths for typical and mean correlations.  Using the SUSY
  technique, mean correlators are obtained as quantum mechanical
  expectation values for an U(2$\vert$1,1) ``superspin''.  In the
  scaling limit, this quantum mechanics is closely related to a
  $0+1$-dimensional Liouville theory, allowing an interpretation of
  the results in terms of simple properties of the zero-energy
  wavefunctions.  Our primary results are the exact two-parameter
  scaling functions for the mean single-particle Green's functions.
  We also show how the Liouville quantum mechanics approach can be
  extended to obtain the full set of multifractal scaling exponents
  $\tau(q)$, $y(q)$ at criticality.  A thorough understanding of the
  unusual features of the present theory may be useful in applying
  SUSY to other delocalization transitions.
\end{abstract}
\pacs{PACS:   }

%\twocolumn
\begin{multicols}{2}
%\narrowtext

\section{Introduction} 

Delocalization transitions control the physical behavior of a number
of electronic systems, including dirty semiconductors, metals and
two-dimensional electron gases in the quantum Hall
regime.\cite{Kramer93,Huckestein95}\ In three dimensions, such
critical points occur at the boundary between a diffusive metal and a
localized insulating phase.  In two or fewer dimensions, however, the
metallic state is generally unstable (``weak localization''), so that
delocalization transitions are typically isolated conducting {\sl
  points} separating two localized phases.  The prototypical example
is the transition between plateaus in the integer quantum Hall effect
(IQHE): Within a model of non-interacting electrons, the localization
length in each disorder-broadened Landau band only diverges at one
isolated energy.\cite{Huckestein95}\ Intense experimental effort has
focused on the quantum Hall plateau transitions, and has led to an
unprecedented characterization of the universal scaling behavior.
Indeed, from the experimental point of view, this system probably
provides the best example of random quantum critical behavior.

Theoretically, however, such systems still present a grand puzzle, in
which but a few pieces are in place.  Some analytical progress has
been made for metal-insulator transitions in $d=2+\epsilon$
dimensions.\cite{Schaefer80,Hikami81,Zirnbauer86}\ But in 2d, despite
a set of simple non-interacting Fermion models which describe the IQHE
plateau transition, a controlled analytic treatment is sorely missing.
Efficient numerical methods have been developed to investigate these
models, and provide a number of significant empirical
observations.\cite{Chalker88,Huckestein95}\ 
In both these cases, scaling is manifest in the vicinity of the
critical point, with a diverging localization length $\xi \sim
|M|^{-\nu}$, where $M$ measures the deviation from criticality.  For
the IQHE, $\nu \approx 7/3$.  For distances shorter than $\xi$, the
single-particle electronic wavefunctions ($\psi(x)$) are extended, but
exhibit complex {\it multifractal} scaling\cite{Huckestein95}.  In particular, each
disorder-averaged moment scales with an independent pair of critical
exponents, which we denote as $\tau(q)$ and $y(q)$ for the $q^{th}$
moment:
\begin{equation}
        \left[ |\psi(x)\psi(0)|^q \right]_{\rm ens.} \sim
        L^{-d-\tau(q)} |x|^{-y(q)}.
\end{equation}
Here the square brackets denote an ensemble average over disorder
configurations, $L$ is the linear extent of the system and $|x|\ll L$
is assumed.  The scaling for essentially all physical quantities can
be formulated in terms of the set of exponents $\nu, \tau(q),$ and
$y(q)$ (a simple example is given in Ref.~\onlinecite{Chalker88a}).

Supersymmetric (SUSY) techniques offer the tantalizing possibility of
a field-theoretic treatment of such delocalization transitions.  SUSY
has a long history in disordered electronic systems, where it was
first introduced by Efetov to describe diffusive
metals.\cite{Efetov83}\ The SUSY non-linear sigma model, when
linearized, provides a {\sl Gaussian or free} field description of a
diffusive metal.  Expansions about the metallic phase in
$d=2+\epsilon$ dimensions give a fixed point which can be extrapolated
to describe a 3d metal-insulator transition.  However, isolated
delocalization transitions in $d \le 2$, such as in the IQHE, do not
afford the luxury of expanding about a diffusive metallic phase.
Recently, Zirnbauer and others have used SUSY to map the
Chalker-Coddington network model for the IQHE transition into an {\sl
  interacting} 1+1-dimensional field theory -- a supersymmetric
antiferromagnetic spin chain.\cite{Zirnbauer94,Lee94,Kondev96}\ 
Unfortunately, this model has resisted all attempts at an analytic
treatment, despite the essentially complete solution of a related
supersymmetric {\sl ferromagnetic} chain which describes transport in
a dirty 2d {\sl chiral} metal.\cite{Balents97}\ Falco and Efetov have
recently applied the SUSY non-linear sigma model in 2d to extract
multifractal wavefunction correlations, but in a crossover regime
rather than at an isolated delocalization transition.\cite{Falko95}\ 

Some analytic progress has been made using a toy model of Dirac
fermions in a random vector potential (RVP), which exhibits a 2d
delocalization transition.\cite{Ludwig94,Mudry96,Chamon96,Kogan96}\ 
This model has the simplifying feature that a zero energy (critical)
wavefunction can be found exactly for any realization of the disorder,
which enables analytic study of wavefunction multifractality.  A
number of different formulations are possible, but a particularly
intriguing approach involves mapping to 2d Liouville field
theory,\cite{Kogan96}\ which has been extensively studied in string
theory.  Away from criticality, however, far fewer results are known,
and at present SUSY techniques have not been successful in this
regard.

In this paper, we study a one-dimensional tight-binding system of
spinless fermions with random hopping matrix elements,  
\begin{equation}
  {\cal H} = - \sum_n  t_n (c_n^\dagger c_{n+1} + c_{n+1}^\dagger c_n).
  \label{FF:Hamiltonian}
\end{equation}
Here the $c's$ are canonical Fermion operators satisfying $\{
c_m,c_n^\dagger \} = \delta_{mn}$, and the random hopping strengths
$t_n$ can be taken positive without loss of generality.  The continuum
limit of this model is in fact a one-dimensional analog of the
2d RVP theory, and many of the same properties obtain.  An exact zero
energy wavefunction is known for each realization of the disorder.
Critical singularities are present in the single-particle density of
states as in the 2d RVP model (but in contrast to the IQHE
transition).  Indeed, for the 1d random hopping model the density of
states {\it diverges} at the band center.

There has been considerable prior work on the 1d transition in the
random hopping model, primarily focussing on properties derivable from
the mean local Green's function: the mean density of states and the
typical localization length.  This work was recently
summarized in Ref.~\onlinecite{McKenzie96}.  Employing a real-space RG
method, D. S. Fisher (DSF) has obtained the spatial dependence of mean
spin-correlation functions in several closely related 1d models: the
McCoy-Wu random transvere field Ising model and random Heisenberg and
XX spin chains.\cite{DFisher94,Fisher95}\ In this paper, we extend the
above analyses using SUSY methods to obtain the spatial dependence of
the {\sl exact} critical and off-critical scaling functions for the
mean Fermion Green's function, summarized in
Eqs.~(\ref{G_omega_scaling}--\ref{last_scalingfn}).  (Unpublished work
by DSF using real-space methods\cite{DSFunpub}\ corroborates our results).  An
important feature not present in the local properties is the emergence
of a {\it mean} localization length which controls the spatial decay
of the average Green's function,
\begin{equation}
  \xi_\epsilon \sim |\ln \epsilon|^2 ,
\end{equation}
with $\epsilon$ the energy from the center of the band.  This length
is much {\it longer} than the {\it typical} localization length
$\tilde\xi_\epsilon \sim |\ln\epsilon|$, found previously by many
authors.  This important distinction between typical and mean
correlation lengths has been emphasized by D. S. Fisher in his
analysis of 1d random spin-chains.

In contrast to the IQHE transition which is mapped into a SUSY
spin-chain, the SUSY formulation of the 1d random hopping model is
equivalent to the quantum mechanics of a {\it single} superspin, with
``Hamiltonian"
\begin{equation}
  H = 2\omega {\cal J}^z + 2m_0 {\cal J}^x - 4g \left( {\cal J}^x \right)^2.
  \label{Hspin:copy}
\end{equation}
The superspin operators $\bbox{\cal J}$ are defined in section III.
This simplification enables us to systematically carry through the
analysis from start to finish to obtain the {\it exact} critical and
off-critical scaling functions.  A key motivation for doing this was
to investigate in detail the novel elements which arise in a SUSY
formulation of a delocalization transition.  Indeed, the ``spin''
Hamiltonian in Eq.~(\ref{Hspin:copy}) has many unconventional
properties.  It is non-Hermitian, requiring a distinction between left
and right eigenstates.  The spin operators themselves are in fact
elements of the non-compact superalgebra U(2$\vert$1,1).  The
non-compactness of the SU(1,1) subalgebra is manifest in the
representations of the spin operators which are infinite-dimensional
(i.e. the ladder of discrete ${\cal J}^z$ eigenstates is infinite).
As demonstrated in sections IV-V, near criticality the system explores
the far reaches of this manifold of spin states, in a manner which can
be described by {\sl Liouville quantum mechanics}, which was recently
introduced in studies of the zero-energy eigenstates in
Ref.~\onlinecite{Shelton97}.  Furthermore, the SUSY Hamiltonian is
{\sl defective}, i.e. the (right) eigenstates do not span the Hilbert
space.  To surmount this difficulty requires the definition of
``pseudo-eigenstates'' to complete the eigenbasis.  A knowledge and
familiarity of these features will likely be crucial to the success of
future work applying SUSY to other (e.g. 2d) critical points.

As a major function of this paper is pedagogy, we have attempted to
present the material in enough detail to allow the reader to
appreciate the technical elements of the calculations. Section II
discusses the model, its relations to various random spin chains, the
continuum limit, and the relevant single-particle Green's functions.
In section III, we describe the mapping to quantum mechanics,
derive the SUSY Hamiltonian and its (super)symmetries, and detail the
organization of states into superspin ladders and super-multiplets.
The exact ground state and a class of excited states needed to compute
the desired correlators are found in sections IV and V, leading to the
final results in Eqs.~(\ref{G_omega_scaling}--\ref{last_scalingfn}).
Lastly, in section VI we pursue the Liouville quantum mechanics
formulation, extending the treatment of Ref.~\onlinecite{Shelton97}\ to
determine the full set of multifractal exponents
\begin{equation}
  \tau(q) = 0, \qquad y(q) = 3/2.
\end{equation}

\section{The Model and Symmetries}

\subsection{Lattice model and continuum limit}

We begin with the free-fermion model, Eq.~(\ref{FF:Hamiltonian}).  We
assume that $t_n$ consists of a large uniform part, $t$, and a small
random piece, $\delta t_n$.  Under a Jordan Wigner transformation,
this model is equivalent to a random exchange spin-$1/2$ XX chain,
\begin{equation}
  {\cal H}_{\rm XX} =  - \sum_n  2t_n ( S^x_n S^x_{n+1} + S^y_n
  S^y_{n+1})  ,
  \label{XX:hamiltonian}
\end{equation}
where $\bbox{S}_n = \bbox{\sigma}_n/2$, with $\bbox{\sigma}$ the usual 
vector of Pauli matrices.

For uniform hopping the single particle states are plane waves, and
${\cal H}$ describes a band at half filling, with zero Fermi energy,
and two Fermi points at $k_{F\pm} = \pm \pi/2$.  With a small random
component in the hopping strengths, the single particle states will be
localized away from the band center, but due to a special particle
hole symmetry (see below) the localization length diverges upon
approaching zero energy.  The density of states is also singular at
zero energy.  To study this delocalization transition, it suffices to
focus on states near zero energy.  Provided $\delta t_n$ is small
compared to the band width $t$, it is legitimate to take a continuum
limit, retaining a narrow shell of pure energy levels near the two
Fermi points.

To this end, we decompose the Fermi fields
as  
\begin{equation}
c_n = (i)^n \psi_{R}(n) + (-i)^n \psi_{L}(n)  ,
\label{lattcont}
\end{equation}
and assume that $\psi$ varies slowly with $n$.
To take the continuum limit we replace $n$ by a continuous
coordinate $x$, and approximate discrete differences
with $x$-derivatives.
For the Fermion hopping term this gives,
\begin{eqnarray}
c_n^\dagger c_{n+1} + c_{n+1}^\dagger c_n & \sim & i\bigg[
\psi_{R}^\dagger \partial_x \psi_R - \left(\partial_x \psi_R^\dagger\right) \psi_R
- (R \leftrightarrow L)\bigg] \nonumber \\
& & -2i (-1)^n \bigg[ \psi_R^\dagger \psi_L - \psi_L^\dagger \psi_R \bigg].
\label{hopping}
\end{eqnarray}
For uniform hopping, $t_n = t \rightarrow dx/2$, the second term is
rapidly varying and can be ignored, giving
the expected (pure) Hamiltonian,
\begin{equation}
{\cal H}_0 = - \int dx [ \psi_R^\dagger i \partial_x \psi_R - \psi_L^\dagger
i \partial_x \psi_L ]  ,
\end{equation}
which describes right and left moving modes at the two Fermi points.

A small random hopping $\delta t_n$ causes scattering between
the plane wave states.
The important Fourier components of $\delta t_n$ are
at $\pi$, since these cause backscattering between
the right and left movers.  We thus decompose
$\delta t_n \rightarrow (-1)^n m(x) dx/2$,
where $m(x)$ is assumed slowly varying.
$ $From the second term in (\ref{hopping}), this leads
to a (random) backscattering term in the
continuum limit:
\begin{equation}
  {\cal H}_1 = -i \int \!dx \, m(x) \left( \psi_R^\dagger \psi_L -
    \psi_L^\dagger \psi_R \right)  .
\end{equation}
Employing a spinor notation, $\psi = (\psi_R, \psi_L)$,
the full continuum Hamiltonian, ${\cal H}_c={\cal H}_0 + {\cal H}_1$ takes 
the form 
\begin{equation}
  {\cal H}_c = \int \! dx \, \psi^\dagger h \psi ,
\label{Hamiltonian}
\end{equation}
with a single-particle Hamiltonian
\begin{equation}
  h = - i \sigma^z \partial_x + m(x) \sigma^y.
\label{hamiltonian} 
\end{equation}

It is convenient to decompose the function
$m(x)$ into a uniform and random piece as
\begin{equation}
  m(x) = m_0 + \tilde{m}(x)  ,
\end{equation}
where $[\tilde{m}]_{\rm ens.} = 0$, with the square brackets denoting
an ensemble average.  Non-zero $m_0$ correspondes to a (uniform)
staggering in the hopping, $\delta t_n \sim (-1)^n m_0$, which opens a
gap in the pure spectrum about the band center.  In the XX spin-chain,
non-zero $m_0$ corresponds to a dimerization in the bond strengths,
and the gap is a spin-gap due to singlet formation across the stronger
bonds.  Both $m_0$ and the energy $\epsilon$ are tuning parameters
which take one away from the delocalized critical point.

\subsection{Symmetries and delocalization}

The lattice free Fermion Hamiltonian Eq.~(\ref{FF:Hamiltonian})
is invariant under the canonical
transformation,
\begin{equation}
c_n \rightarrow (-1)^n c_n^\dagger  ,
\label{phsymm}
\end{equation}
due to time-reversal and particle-hole symmetries,
present even with random hopping strengths.
As a consequence of this symmetry,
the single-particle wave functions can be chosen real
(time reversal invariance) and come in conjugate pairs
with energy $\pm \epsilon$.  Specifically, for a given eigenfunction,
$\phi_\epsilon(n)$ at energy $\epsilon$, there is a partner eigenstate
with energy $-\epsilon$, given by $\phi_{-\epsilon} = (-1)^n \phi_\epsilon(n)$.
At zero energy one thus anticipates special properties, as discussed below.

In the continuum, the symmetry Eq.~(\ref{phsymm})
becomes an invariance of ${\cal H}_c$ under the
canonical transformation,
\begin{equation}
\psi_\alpha \rightarrow \psi_\alpha^\dagger  ,
\end{equation}
for $\alpha = R,L$.  This symmetry restricts the allowed form
of the single particle Hamiltonian, $h$.
Specifically, $h$ cannot have terms (with no gradients) proportional
to $\sigma^x, \sigma^z$ or the identity.  
A generic random 1d tight binding Fermion model,
in which the density of states is regular
and all the eigenstates are localized,
would {\it not}
be particle-hole symmetric, and additional terms, such as a spatially
random $\sigma^x$ term, would be present in the continuum
Hamiltonian.  The above symmetry is clearly crucial for the existence
of delocalization at the band center, $\epsilon=0$.

An (un-normalized) extended state at zero energy can in fact be directly extracted
from the continuum wave equation: $h \Phi(x) =0$,
where $\Phi$ is a two-component wave function.
Writing 
\begin{equation}
  \Phi(x) = \phi_\pm(x) \pmatrix{1\cr \pm1 \cr},
\end{equation}
the scalar function $\phi(x)$ satisifies,
\begin{equation}
  (\partial_x \pm m(x) ) \phi_\pm = 0.
\end{equation}
This can be integrated to give
\begin{equation}
  \phi_\pm (x) \propto e^{\pm \int^x dx^\prime m(x^\prime)}  .
  \label{wavefunction}
\end{equation}
For random $m(x)$, with mean zero, this wavefunction is clearly {\it
  not} exponentially localized.  If the random function $m(x)$ has
short-ranged spatial correlations, the logarithm of the wave function
undergoes a 1d random walk.  For a Gaussian distribution of $m(x)$ the
(unnormalized) wavefunction is log-normally distributed.  This wave
function is a one-dimensional analog of the exact zero energy
wavefunctions written down for 2d free Fermions described by a Dirac
theory with random vector potential.  As in the 2d case, the
wavefunction is very broadly distributed, and it's correlations can be
characterized by a multi-fractal scaling description.  We return to a
discussion of the multi-fractal characteristics of this wave function
in Section~\ref{sec:multifractal}, where we compute the multi-fractal
spectrum explicitly, following recent work by Shelton and
Tsvelik.\cite{Shelton97}\ 

Away from criticality, for non-zero
$m_0$,
the zero energy wavefunctions
in Eq.~(\ref{wavefunction})  are exponentially growing
and decaying functions, $\phi_\pm(x) \sim e^{\pm m_0 x}$.  While they
are non-normalizable in infinite space, for a finite system they
describe solutions which decay exponentially into the system,
with an associated localization (or correlation) length, 
$\tilde{\xi} = 1/m_0$.
The critical exponent,
defined via
\begin{equation}
  \tilde{\xi} \sim m_0^{-\tilde{\nu}}   ,
\end{equation}
is $\tilde{\nu} =1$.  As emphasized by DSF, in addition to the length
$\tilde{\xi}$ which describes the decay of a typical (unaveraged)
correlation function, there is another divergent length, $\xi$, which
describes the decay of ensemble averaged correlation functions.
Consistent with arguments by DSF, we find below that this latter
length diverges more rapidly with an exponent $\nu=2$.

Two similar lengths may be defined by approaching the
critical point at finite energy $\epsilon$, but with zero mass $m_0
=0$.  Using the Thouless construction from the local Green's function,
previous authors have found a (typical) localization length
$\tilde\xi_{\epsilon} \sim |\ln\epsilon|$.  In constrast,
employing a real-space RG approach, DSF has shown that
mean correlation functions decay with a longer length,
which varies as
$\xi_\epsilon \sim |\ln\epsilon|^2$.

Another important characteristic of the above exact
zero energy wavefunction is that it is {\it nodeless}, for
each and every realization of the random potential $m(x)$.  Because of
this, critical properties of the 1d localization transition at
$\epsilon = 0$ are contained in the 
ensemble averaged single-particle Green's function,
in contrast to the 
conventional Anderson transition
in higher dimensions.

Below we briefly consider symmetry properties of the single
Fermion Green's function, and obtain expressions
in the continuum limit.  The next sections are devoted
to evaluating the ensemble averaged Green's function
using supersymmetry methods.

\subsection{Green's Functions}

Consider the single Fermion Green's function
at energy $\epsilon$, defined as
\begin{equation}
{\cal G}(n,n^\prime;\epsilon+i\omega) =
i \int_0^\infty dt e^{i(\epsilon+i\omega)t} \langle v| c_n(t)
c_{n^\prime}^\dagger (0) |v \rangle  ,
\label{green}
\end{equation}
where $|v \rangle$ denotes the Fermion vacuum, and
$c(t) = e^{i{\cal H}t} c e^{-i{\cal H}t}$ with
${\cal H}$ the lattice Hamiltonian.  
Here $\omega$ is a small imaginary part to the energy.
In practice below, we will take the
real part of the energy to be zero, $\epsilon = 0$,
calculate ${\cal G}(i\omega)$ for real $\omega$,
and then extract the energy dependence via an analytic continuation.

The spectral decomposition of ${\cal G}$ in terms of the
exact eigenstates,
$\phi_{\epsilon}$,
takes the form
\begin{equation}
{\cal G}(n,n^\prime;i\omega) = \sum_\epsilon { {\phi_{\epsilon}(n) 
\phi_\epsilon(n^\prime) } \over {\epsilon - i \omega}}  .
\label{spectral}
\end{equation}
Using the symmetry property $\phi_{-\epsilon}(n) = (-1)^n
\phi_\epsilon(n)$, one can readily show that ${\cal G}(n,n^\prime,
i\omega)$ is {\it real} and even in $\omega$ for $n-n^\prime$ odd, and
purely {\it imaginary} and odd in $\omega$ for $n-n^\prime$ even.  For
example, with $n-n^\prime$ even, Eq.~(\ref{spectral}) can be
rewritten as
\begin{equation}
{\cal G}(n,n^\prime;i\omega) = \sum_{\epsilon >0} \phi_{\epsilon}(n) 
\phi_\epsilon(n^\prime)  {{-2i \omega} \over {\epsilon^2 + \omega^2}}  .
\end{equation}

The Green's function, as defined in Eq.~(\ref{green}),
can be re-expressed in terms of the continuum
Fermion fields, $\psi(x)$, by employing the decomposition Eq.~(\ref{lattcont}).
For $n-n^\prime \gg 1$, the discrete separation can be replaced by
a continuous distance
$ x-x^\prime$ (lattice constant equal to one).
One finds
\begin{equation}
  {\cal G}(n,n^\prime;i\omega) = i^{n-n^\prime}  \sum_{\alpha , \beta}
  (-1)^{\alpha n + \beta n^\prime} G_{\alpha \beta} (x,x^\prime;i\omega)  ,
  \label{Glattcont}
\end{equation}
with $\alpha$ and $\beta$ running over the
two spinor components, denoted
as either
(R,L),  $(0,1)$ or $(\uparrow, \downarrow)$.
Here $G_{\alpha \beta}$ is defined in terms of
the continuum Fermion fields, $\psi$, as
\begin{equation}
G_{\alpha \beta}(x,x^\prime; i\omega) = i \int_0^\infty dt e^{-\omega t}
\langle v| \psi_\alpha(x,t) \psi_\beta^\dagger(x^\prime,0) |v\rangle ,
\end{equation}
with $\psi(t) = e^{i{\cal H}_ct} \psi e^{-i{\cal H}_ct}$ and
${\cal H}_c$ the {\it continuum} Hamiltonian.  
This continuum Green's function
can alternatively be expressed in terms of the
single-particle Hamiltonian, $h$ in Eq.~(\ref{hamiltonian}),
as
\begin{equation}
G_{\alpha\beta}(x,x^\prime; i\omega) = \langle x,\alpha| {1 \over {h - 
i \omega}} |x^\prime,\beta \rangle  ,
\end{equation}
where $|x,\alpha \rangle$ denotes a Fermion at position $x$
with ``spin"-component $\alpha$.

Of interest is the behavior of the
single-particle Green's function upon ensemble averaging over
disorder realizations.  To be concrete,
we take the random function $\tilde{m}(x)$
to be Gaussian with,
\begin{equation}
  [ \tilde{m}(x) \tilde{m}(x^\prime) ]_{\rm ens.} = 2g \delta(x-x^\prime)  .
\end{equation}
Ensemble averaged Green's functions,
which we denote with an overbar, become translationally
invariant:
\begin{equation}
\overline{{\cal G}}(x;i\omega) = {1 \over N} \sum_{n=1}^N 
[{\cal G}(n+x,n;i\omega) ]_{\rm ens.}  ,
\end{equation}
with $N \rightarrow \infty$ the number of sites in the tight binding
lattice.  From Eq.~(\ref{Glattcont}), this is related to the averaged continuum
Green's functions,
\begin{equation}
  \overline{G}_{\alpha \beta}(x;i\omega)  = 
  [G_{\alpha \beta}(x,0;i\omega)]_{\rm ens.},
\end{equation}
via
\begin{equation}
  \overline{{\cal G}}(x;i\omega)  = i^x \sum_\alpha (-1)^{\alpha x}
  \overline{G}_{\alpha \alpha}(x;i\omega)   ,
\end{equation}
with the separation $x$ either even or odd.

The mean density of states for the original lattice Fermions
can be written,
\begin{equation}
\rho(\epsilon) = \lim_{\omega \rightarrow 0} {1 \over \pi} {\rm Im} \overline{{\cal G}}(x=0;\epsilon + i\omega)  .
\label{DOS}
\end{equation}
We shall also be interested in the spatial dependence of
the correlation function,
\begin{equation}
C(x,\epsilon) =  {1 \over \pi} {\rm Im} \overline{{\cal G}}(x;\epsilon + i 0^+) .
\label{C_x}
\end{equation}
This function is expected to decay
exponentially with a mean correlation length $\xi$.
Due to the delocalized zero energy wave function
(at $m_0=0$),
$\xi(\epsilon)$ should diverge upon approaching
the band center, $\epsilon \rightarrow 0$.

In Section III  we will construct a generating functional
which can be used to extract $\overline{G}$.
Our strategy will be to calculate $\overline{G}(x,i\omega)$ for {\it real}
$\omega$, and then perform an analytic continuation to extract
the density of states and $C(x,\epsilon)$.

\subsection{Related Random models}

As noted above, under a Jordan-Wigner transformation the lattice
free-Fermion Hamiltonian ${\cal H}$ in Eq.~(\ref{FF:Hamiltonian}) is
identical to a random exchange spin-1/2 XX chain.  Some properties of
the spin-chain can be extracted from the Fermion density of states,
specifically the specific heat and the z-component magnetization,
$\langle S^z \rangle$, in response to a magnetic field along the
z-axis.  Unfortunately, spin correlation functions are notoriously
difficult to extract from the Free-fermion representation, due to the
non-local relation between spin and Fermion operators (Jordan-Wigner
string).  Nevertheless, one expects that the correlation decay length
of the Fermion Green's functions will also control the decay of spin
correlations.

As shown originally by Shankar and Murthy, a second spin model which
is equivalent upon fermionization to the free Fermion model ${\cal
  H}$, is the 1d random quantum Ising chain in transverse field, with
Hamiltonian
\begin{equation}
  {\cal H}_{\rm I} = \sum_n \left[ 2K_{1,n} S_n^x + 4 K_{2,n} S_n^z
    S_{n+1}^z \right] .
   \label{I:hamiltonian}
\end{equation}
Here $K_1$ and $K_2$ are spatially random field and Ising exchange
constants, respectively.  This model is an anisotropic
(``time-continuum") version of a 2d classical Ising model
with random exchange interactions perfectly correlated
in one of the two directions, a model
first studied by McCoy and Wu.
For completeness, we sketch this fermionization procedure
in Appendix A, where we show that the low energy properties
of ${\cal H}_{\rm I}$ follow from the properties
of the single-particle Hamiltonian $h$ in Eq.~(\ref{hamiltonian}).

Finally, we should mention the equivalence between the
free Fermion Hamiltonian ${\cal H}$ and a 1d model
of quantum particles connected by random strength harmonic springs,
a model first introduced and analyzed by Dyson almost 50 years
ago.\cite{Dyson}\

\section{Quantum mechanics}

As discussed above, much of the information of interest is contained
in the mean single-particle Green's function
$\overline{G}_{\alpha\beta}(x,\epsilon+i\omega)$.  In the following we will
construct a generating functional which can be used to extract
$\overline{G}$.  Our strategy will be to calculate
$\overline{G}(x,i\omega)$ for {\it real} $\omega$, and then perform an
analytic continuation to extract the density of states and
$C(x,\epsilon)$.

\subsection{SUSY generating functional}

Our analysis is based on employing the well-known
field-theoretic representation of an operator inverse
to express the ensemble averaged Green's function,
\begin{equation}
   \overline{G}_{\alpha \beta}(x,i\omega) =  [<x,\alpha| {1 \over {h - 
   i\omega}} |0,\beta>]_{\rm ens.} ,
   \label{Greens}
\end{equation}
as a functional integral,
\begin{equation}
   \overline{G}_{\alpha \beta}(x, i\omega) = i \langle \psi_\alpha (x) 
   \overline{\psi}_\beta (0) \rangle_S  .
   \label{average}
\end{equation}
with
\begin{equation}
  \langle O \rangle_{S} = \left[\int D\psi D\overline{\psi} D\xi
    D\xi^* O  e^{-S}\right]_{\rm ens.}.
  \label{stat:avg}
\end{equation}
Here the functional integration is over Grassmann
fields $\psi(x)$ and complex fields $\xi(x)$, with the action
\begin{equation}
  S = \int dx [ \overline{\psi}(ih + \omega) \psi + \xi^* (ih + \omega) \xi ] .
  \label{action1}
\end{equation}
The most noteworthy point is that in Eq.~(\ref{stat:avg}), we have 
included the complex scalar field $\xi$ in order to cancel the 
Fermionic determinant which naturally occurs due to the $\psi$ 
integration.  In doing so, we obtain the supersymmetric form in 
Eq.~(\ref{action1}).  This enables an ensemble average over
the Gaussian disorder $\tilde{m}(x)$ to be readily performed.

For ease in presentation, it is convenient to 
speak of a slightly simpler object, the partition function
\begin{equation}
{\cal Z} = \int D\psi D\overline{\psi} D\xi D\xi^* e^{-S},
\end{equation}
where the functional integration is, as before, over Grassmann fields 
$\psi(x)$ and complex fields $\xi(x)$, with the action of 
Eq.~(\ref{action1}).  Correlation functions are obtained by simply 
inserting the appropriate fields after the integration measure and 
ensemble averaging.  The 
crucial cancellation of the Fermionic and Bosonic determinants then 
gives the trivial identity $\langle 1\rangle = 1$, or ${\cal Z} = 1$.  
$ $From the functional integral formulation, the reason for generating 
Green's functions for real $\omega$ becomes clear: while the Fermionic 
functional integral is always well defined, the bosonic one is only 
convergent provided that the action is bounded below.  This is the 
case here provided $\omega > 0$, since $h$ is hermitian.

\subsection{Tranformation to quantum formulation}

The above action corresponds to a random one-dimensional statistical
mechanics problem.  After ensemble averaging over the random function
$m(x)$, the model is translationally invariant.  Our approach is to
extract the transfer matrix, $\hat{T} = e^{-H}$, which can be used to
reconstruct the averaged generating function:
\begin{equation}
  [ {\cal Z}]_{\rm ens.} = \lim_{L \rightarrow \infty} {\rm STr} e^{-LH} ,
\end{equation}
where $L$ is the length of the system.  The symbol STr indicates the
supertrace, defined by
\begin{equation}
  {\rm STr} O = {\rm Tr} \left[ (-1)^{N_f} O \right],
\end{equation}
where Tr is the conventional trace, and $N_f$ is a Fermion number
operator defined below.  Although $H$ is an operator, easily
expressed in terms of Fermi and Bose operators (see below), it is
``zero-dimensional," being independent of the spatial coordinate $x$.
The problem is thus reduced to studying the quantum mechanics of the
``Hamiltonian" $H$.  As usual, the spectrum of $H$ contains
information about the correlation length of the 1d system - in this
case the localization length.

To extract the operator $H$, we massage the action Eq.~(\ref{action1})
into the form of a coherent state path integral with
$x$ playing the role of imaginary time.  To this end
we let $\xi_\downarrow \rightarrow
-\xi^*_\downarrow$, leaving $\xi_\uparrow$ unchanged.
Similarly we transform
the independent Grassmann fields as
$\psi_\downarrow \rightarrow - \overline{\psi}_\downarrow$
and $\overline{\psi}_\downarrow \rightarrow \psi_\downarrow$,
leaving the spin-up Grassmann fields unchanged.
The action can then be written
\begin{equation}
S = \int dx  ({\cal L}_0 + {\cal L}_\omega + {\cal L}_m )   ,
\end{equation}
with
\begin{equation}
{\cal L}_0  = \overline{\psi} \partial_x \psi + \xi^* \partial_x \xi  ,
\end{equation}
\begin{equation}
{\cal L}_\omega = \omega ( \overline{\psi} \psi + \xi^* \xi ),
\end{equation}
and ${\cal L}_m = m(x) A$ with
\begin{equation}
A = \overline{\psi}_\uparrow   \overline{\psi}_\downarrow
- \psi_\uparrow \psi_\downarrow + (\psi \rightarrow \xi)
  .
\end{equation}
Notice that ${\cal L}_0$ is now in the standard form for a coherent
state path integral if $x$ is reinterpreted as an imaginary time coordinate.

Before extracting $H$, we perform an ensemble average over the
disorder.  Since we have assumed a Gaussian distribution, this is
readily performed to extract $[{\cal Z}]_{\rm ens.}$.  The only term
in the action which is modified is ${\cal L}_m$, which now becomes,
\begin{equation}
{\cal L}_m^{\rm ens.} = m_0 A - g A^2  .
\end{equation}

The transfer ``Hamiltonian" $H$ can now be read off, since the full
action takes the form, $S = \int_x ({\cal L}_0 + H(\psi,\xi))$.  In
passing to the Hamiltonian framework, the Grassmann fields are
replaced by Fermion operators, $\psi \rightarrow f$, $\overline{\psi}
\rightarrow f^\dagger$, and the complex fields by Bose operators, $\xi
\rightarrow b$, $\xi^* \rightarrow b^\dagger$, where $f$ and $b$
satisfy canonical commutation relations:
\begin{equation}
[f_\alpha,f^\dagger_\beta]_- = [b_\alpha,b^\dagger_\beta] = \delta_{\alpha\beta}.
\end{equation}
The resulting ``Hamiltonian" is,
\begin{equation}
H = \omega[f^\dagger f + b^\dagger b ] + m_0 A - g A^2 ,
\label{transferH}
\end{equation}
with
\begin{equation}
A = f^\dagger_\uparrow f^\dagger_\downarrow - f_\uparrow f_\downarrow + (f \rightarrow b)    .
\end{equation}
Although we will hereafter refer to $H$ as a Hamiltonian,
it is important to keep in mind that this operator
is {\it not} Hermitian.

Since $H$ does not conserve the Fermion number, $N_f = f^\dagger f$,
it is convenient to perform a particle hole transformation,
defining new Fermion fields via a canonical transformation
\begin{equation}
F_\uparrow = f_\uparrow , \quad F_\downarrow = f_\downarrow^\dagger ,
\end{equation}
where $F^\dagger F$ commutes with $H$.  To preserve the
Bose-Fermi supersymmetry one can also define,
\begin{equation}
B_\uparrow = b_\uparrow , \quad B_\downarrow = b_\downarrow^\dagger .
\end{equation}
However, $B_\downarrow$ does {\it not} satisfy the canonical
Boson commutator, but rather, $[B_\downarrow^{\vphantom{\dagger}},
B_\downarrow^\dagger] = -1$.  To restore the canonical form we define
\begin{equation}
\overline{B} = B^\dagger \sigma^z ,
\end{equation}
so that $B$ and $\overline{B}$ satisfy,
\begin{equation}
[B_{\alpha}, \overline{B}_{\beta} ] = 
\delta_{\alpha
\beta}  .
\end{equation}
However, it must be kept in mind that $\overline{B} \ne B^\dagger$.

In term of these new operators, the
Hamiltonian $H$ becomes
\begin{equation}
H = \omega[F^\dagger \sigma^z F + \overline{B} \sigma^z B] + m_0 A - g A^2 ,
\end{equation}
with 
\begin{equation}
A = F^\dagger \sigma^x F + \overline{B} \sigma^x B .
\end{equation}

At this stage it is convenient to express the Green's function
Eq.~(\ref{Greens}) as a supertrace over quantum states.  Consider the
representation Eq.~(\ref{average}) in terms of Fermions.  After
transforming to the coherent state path integral form, the Grassmann
fields can be replaced by Fermion operators: $\psi(x) \rightarrow
e^{xH} f e^{-xH}$.  The average in Eq.~(\ref{average}) is replaced by
a supertrace over quantum states:
\begin{equation}
  \langle O(\psi,\xi) \rangle_S \rightarrow \langle O(f,b) \rangle
  \equiv {\rm STr}[O e^{-LH}] .
\end{equation}
This gives,
\begin{equation}
  \overline{G}_{\alpha \beta}(x;i\omega) = i(-1)^\alpha  {\rm
    STr}[F_\alpha(x) F^\dagger_\beta e^{-LH} ]  , 
\label{GSTr}
\end{equation}
with $F_\alpha(x) = e^{xH} F_\alpha e^{-xH}$.

\subsection{Supersymmetry}

In order to disucss the symmetries of the Hamiltonian $H$ it is useful
to introduce a four-component superfield,
\begin{equation}
\Psi = (F, B), \; \; \overline{\Psi} = 
(F^\dagger,\overline{B}).
\end{equation}
We will use latin indices ($a,b,...$) to denote the
Fermion/Boson label, i.e. $a=B,F \leftrightarrow 0,1$.
$ $From this superfield,
one may build a three-component superspin,
\begin{equation}
\bbox{\cal J}_{ab} = {1 \over 2} \overline\Psi_{a\alpha}  \bbox{\sigma}_{\alpha\beta} \Psi_{b\beta} ,
\end{equation}
where sums on the greek indices are implied.
One can also define a set of ``charges",
\begin{equation}
Q_{ab} = \overline{\Psi}_{a\alpha} \Psi_{b\alpha}  ,
\label{charges}
\end{equation}
(sum on $\alpha$).

The diagonal components
of $\bbox{\cal J}$ have special significance.  In the Fermionic
sector,
\begin{equation}
\bbox{\cal J}_{11} \equiv \bbox{S} = F^\dagger {\bbox{\sigma} \over 2} F
\end{equation}
forms a set of ordinary Hermitian SU(2) spin operators, satisfying
$[S^i,S^j] = i\epsilon^{ijk}S^k$.  In the bosonic sector, we may
similarly define three other currents,
\begin{equation}
\bbox{\cal J}_{00} \equiv \bbox{J} = \overline{B} {\bbox{\sigma} \over 2} B.
\end{equation}
These also satisfy the SU(2) algebra, $[J^i,J^j] = i\epsilon^{ijk}
J^k$.
However, although $J^z$ is Hermitian, $J^x$ and $J^y$ are
{\it anti}-Hermitian.  
One could define
a Hermitian set of operators,
multiplying the $x$ and $y$ components of $J$ by a factor of $i$.
These would then satisfy SU(1,1) commutation relations,
instead of
SU(2).

A useful object is the total spin-current
\begin{equation}
\bbox{\cal J} = \bbox{\cal J}_{aa},
\end{equation}
where again the sum on repeated indices is implied.
This spin-current commutes with the
charges:
\begin{equation}
[\bbox{\cal J}, Q_{ab}] = 0.
\label{conserved_currents}
\end{equation}
Since the
Hamiltonian can be expressed
in terms of this total spin-current,
\begin{equation}
H = 2\omega {\cal J}^z + 2m_0 {\cal J}^x - 4g \left( {\cal J}^x \right)^2,
\label{Hspin}
\end{equation}
the charges $Q_{ab}$ also commute with $H$.  Thus $Q_{ab}$ generate a
set of (super)symmetries of $H$.  Because of their importance, it is
convenient to name them individually:
\begin{eqnarray}
  N_B \equiv Q_{00} = \overline{B} B, & \qquad & N_F \equiv Q_{11} =
  F^\dagger F, \\
  Q \equiv Q_{01} = \overline{B} F, & \qquad & \overline{Q} \equiv
  Q_{10} = F^\dagger B. 
\end{eqnarray}
These latter two operators are Fermionic ``charges", which will be
extremely useful in determining the ground state of $H$.  They obey
\begin{eqnarray}
  Q^2 = \overline{Q}^2 & = & 0, \\
  \{ Q,\overline{Q} \} & = & N,
\end{eqnarray} 
where the total charge is defined as
\begin{equation}
  N = N_B + N_F,
\end{equation}
which commutes with {\sl all} sixteen of the U($2\vert$1,1) currents.

\subsection{Hilbert space and Representations}

Finding the ground state and low energy excitations
of $H$ is complicated by the enormity of the Hilbert space.
Indeed,
since 
the number of bosons with spin $\alpha$,
$n_\alpha = b_\alpha^\dagger b_\alpha$, is unbounded,
the Hilbert space is actually infinite.
One basis of states spanning the Hilbert space
may be written as a direct product
of bosonic and Fermionic states:
\begin{equation}
|n_\uparrow n_\downarrow \alpha \rangle = |n_\uparrow
n_\downarrow\rangle \otimes |\alpha_F\rangle,
\end{equation}
where $n_\uparrow,n_\downarrow = 0...\infty$
are the number of up- and
down-spin bosons, respectively.  The parameter $\alpha_F$ labels the
Fermionic sector, which is spanned by the Fermionic vacuum $|{\rm vac}\rangle$,
which is anihillated by $F$,  
and three other states
\begin{eqnarray}
|\uparrow\rangle & = & F_\uparrow^\dagger |{\rm vac}\rangle, \\
|\downarrow\rangle & = & F_\downarrow^\dagger |{\rm vac}\rangle, \\
|\downarrow\uparrow\rangle & = &
F_\uparrow^\dagger F_\downarrow^\dagger |{\rm vac}\rangle.
\end{eqnarray}

As is usual in a quantum mechanics problem, we may simplify matters by
choosing a maximal set of commuting variables, whose eigenvalues are
good quantum numbers.  In this case, $N_B$ and $N_F$ are obvious
choices.  The Fermionic charges $Q$ and $\overline{Q}$ cannot, of
course, be diagonalized.  They can, however, be combined to form two
additional bosonic charges
\begin{equation}
  \Gamma = \overline{Q} Q, \qquad  \overline{\Gamma} = Q\overline{Q}.
\end{equation}
It is straightforward to show that 
\begin{eqnarray}
  \Gamma^2 = N\Gamma, & \qquad & \overline{\Gamma}^2 = N
  \overline{\Gamma}, \\
 \Gamma + \overline{\Gamma} = N, & \qquad &  \overline{\Gamma} \Gamma = 0.
\end{eqnarray}
These relations imply that the eigenvalues of
$(\Gamma,\overline{\Gamma})$ are either $(0,N)$ or $(N,0)$.
The four operators $N_B, N_F, \Gamma, \overline{\Gamma}$ form the
desired set of good quantum numbers, and it is desirable to reorganize
the states given above into a basis diagonal in these variables.

\subsubsection{SU(2)}

Before proceeding to determine this basis, consider first the
fermionic sector of the theory.  The fermion number is in fact related
to the total spin via
\begin{equation}
  S^2 = s(s+1),
\end{equation}
with
\begin{equation}
  s = N_F(2-N_F)/2.
\end{equation}
So we can think of $N_F$ as determining the representation of SU(2).
Note that the singlet ($s=0$) representation occurs twice -- for $N_F
= 0,2$.

\subsubsection{SU(1,1) and bosonic ladders}

Similarly, the bosonic states may be separated into multiplets
with fixed $N_B = n_\uparrow - n_\downarrow -1$.  Each such multiplet
is actually a distinct irreducible representation of SU(1,1).  To see
this, consider the Casimir operator
\begin{equation}
  J^2 = (J^x)^2 + (J^y)^2 + (J^z)^2 = (N_B^2 + 2 N_B)/4.
\end{equation}
Fixing $N_B$ thus fixes the ``total spin'' of the SU(1,1)
representation.  Following the analogy with SU(2), we may label the
states by their total spin and, e.g. the spin $J^z$ along the z-axis,
\begin{eqnarray}
  J^2 |j n \rangle & = & j(j+1) |j n\rangle, \\
  N_B |j n \rangle & = & 2j |j n \rangle, \\
  J^z |j n \rangle & = & \left[ {{1+|2j+1|} \over 2} + n \right] |j n\rangle.
\end{eqnarray}
Note that in the last equation we have departed from the usual
convention for denoting $J^z$ eigenvalues.  This is convenient
because the quantum number $n$ as we have defined it takes integer
values $n=0,\ldots\infty$.  The total spin can take half integer
values $j=0,\pm 1/2, \pm 1, \ldots$.

It is also helpful to have explicit expressions in terms of the
previous basis.  Since the $j^{\rm th}$ block of states corresponds to
a ladder satisfying $n_\uparrow - n_\downarrow = 2j+1$, it can be
conveniently generated using the raising and lowering operators:
$J^\pm = J^x \pm i J^y$.  As usual, the lowest weight state in each
ladder, denoted $|j 0\rangle$, is constructed to be annihilated by
$J_- = -b_\uparrow b_\downarrow$:
\begin{equation}
|j  0 \rangle = \cases{{1 \over \sqrt{(2j+1)!}}
\left(b_\uparrow^\dagger\right)^{2j+1} |v\rangle & $ j \geq -1/2$ \cr
{1 \over \sqrt{|2j+1|!}}
\left(b_\downarrow^\dagger\right)^{-(2j+1)} |v\rangle & $ j < -1/2$},
\end{equation}
where $|v\rangle$ denotes the bosonic vacuum: $b_\alpha |v\rangle =0$. 
Each ladder is constructed by acting on the associated lowest
weight state with powers of $J^+ = b_\uparrow^\dagger b_\downarrow^\dagger$;
\begin{equation}
|j n\rangle = \left({{|2j+1|!} \over {n!(n+|2j+1|)!}}\right)^{1/2}
\left(J^+\right)^n |j 0\rangle ,
\end{equation}
where $n$ runs from zero to $\infty$.

\subsubsection{SUSY ladders}

Clearly the set of eigenvalues of $N_B,N_F,\Gamma,\overline{\Gamma}$
is insufficient to distinguish all the states in the Hilbert space.
To provide a unique labelling, we will choose to diagonalize the
additional operator ${\cal J}^z = J^z + S^z$.  This choice is natural
in that ${\cal J}^z$ commutes with the other four diagonal charges.
Moreover, the other current present in the Hamiltonian, ${\cal J}^x$,
leaves the original four quantum numbers unchanged, mixing only
different values of ${\cal J}^z$.  The collection of states with
different ${\cal J}^z$ but with the other four charges fixed may be
viewed as the basis for a representation of the algebra of the
$\bbox{\cal J}$ operators, i.e. a peculiar (non-unitary)
representation of SU(2).

Two sets of such states are easily constructed.  These are
\begin{eqnarray}
  |2j, 0, 0, N, (|N+1|+1)/2+n\rangle & = & |j n\rangle \otimes
  |{\rm vac}\rangle, \\ 
  |2j, 2, N, 0, (|N-1|+1)/2+n\rangle & = & |j n\rangle \otimes
  |\downarrow\uparrow\rangle,
\end{eqnarray}
where the quantum numbers inside the bras on the left-hand-side denote
eigenvalues of $(N_B, N_F, \Gamma, \overline{\Gamma}, {\cal J}^z)$, in
that order.

\end{multicols}
States with $N_F=1$ are slightly more complicated, since they can
involve linear combinations of up or down fermions.  For $N \neq 0$,
these are
\begin{eqnarray}
   |2j, 1, (N-|N|)/2, (N+|N|)/2, |N|/2 + n\rangle 
  & =  &   \sqrt{{n+|N|}
     \over {2n+|N|}} |j n\rangle \otimes |\downarrow\rangle 
  + \sqrt{n \over {2n+|N|}} |j n-1\rangle \otimes
  |\uparrow\rangle,   \\
   |2j,1, (N+|N|)/2, (N-|N|)/2, |N|/2 + n \rangle  & = & 
 \sqrt{n \over {2n+|N|}} |j n\rangle \otimes |\downarrow\rangle
  + \sqrt{{n+|N|} \over {2n+|N|}} |j n-1\rangle \otimes
  |\uparrow\rangle, 
\end{eqnarray}
where in the first set above $n=0,1,2,\ldots\infty$, while in the second
set $n=1,2,\ldots\infty$.

Apparently these two ladders of states become identical for $N=0$.
What happens in that case?  There are two states
which are eigenstates of $N_B=-1$,$N_F=1$, and ${\cal J}^z = n$:
\begin{equation}
  |-1/2,n\rangle \otimes |\downarrow\rangle
  ,|-1/2,n-1\rangle\otimes|\uparrow\rangle. 
\end{equation} 
If, however, we attempt to diagonalize, e.g. $\Gamma$, in this
two-dimensional space, we find that it is impossible!  In fact, there
is only a single eigenstate for each $n$,
\begin{equation}
  |-1,1,0,0,n\rangle  = \cases{{1 \over \sqrt{2}}\left[
      |-1/2,n\rangle \otimes 
    |\downarrow\rangle +|-1/2,n-1\rangle\otimes|\uparrow\rangle\right] &
  $n >0$ \cr |-1/2,0\rangle \otimes |\downarrow\rangle & $n=0$ \cr},
\label{zero_ladder}
\end{equation}
which are annihilated by both $Q$ and $\overline{Q}$.
Clearly this set of states does not span the full subspace.  To
complete it, we may define an orthogonal ladder of states
\begin{equation}
  |-1,1,*,*,n\rangle = {1 \over \sqrt{2}}\left[ |-1/2,n\rangle \otimes
      |\downarrow\rangle -|-1/2,n-1\rangle\otimes|\uparrow\rangle\right],
    \label{star_ladder}
\end{equation}
for $n=1,2,\ldots\infty$.  It is important to note that
{\sl $|-1,1,*,*,n\rangle$ is  not an eigenstate of $\Gamma$ and
  $\overline{\Gamma}$}.  Instead, acting with these operators on
$|-1,1,*,*,n\rangle$ gives back $|-1,1,0,0,n\rangle$, i.e. $\Gamma$
and $\overline{\Gamma}$ act like projection operators in this subspace.

\begin{multicols}{2}
\subsection{Eigenstates}

\subsubsection{supermultiplets}

In a conventional quantum mechanics problem, we can look for
eigenstates of the Hamiltonian separately for each distinct set of
eigenvalues of the chosen commuting operators.  
As we have seen above, most of the
states in the Hilbert space can be specified in this way. To avoid the
exceptions for the moment, consider first the sectors with $N\neq 0$.
Then we may find {\sl right} eigenstates of the Hamiltonian
\begin{equation}
  H |N_B,N_F,\Gamma,\overline{\Gamma},E\rangle = E
  |N_B,N_F,\Gamma,\overline{\Gamma},E\rangle , 
\end{equation}
where each such state can be expanded in a basis of appropriate
${\cal J}^z$ eigenstates, i.e.
\begin{equation}
  |N_B,N_F,\Gamma,\overline{\Gamma},E\rangle = \sum_{{\cal J}^z}
  \chi^{N_B,N_F,\Gamma,\overline{\Gamma}}_E({\cal J}^z)
  |N_B,N_F,\Gamma,\overline{\Gamma},{\cal J}^z\rangle.
\end{equation}
In a theory with ordinary bosonic symmetries, completely specifying
the quantum numbers usually determines a unique set of energies --
accidental degeneracies are rare.  In a SUSY theory, however, the
additional fermionic generators $Q$ and $\overline{Q}$ lead to
extra relations between states.  

To see this, note that since $Q$ and $\overline{Q}$ commute with
$H$, acting with them upon an eigenstate must either produce another
eigenstate with the same energy or annihilate it.  Because $Q$ and
$\overline{Q}$ do not commute with the four diagonal operators, however,
they {\sl must} change these eigenvalues, and hence connect distinct
states.  Because the total charge $N$ {\sl does} commute with $Q$ and
$\overline{Q}$, it (along with the energy $E$) can be used to
characterize a multiplet of states connected in this way.
\begin{figure}[hbt]
  \epsfxsize=3.5in\epsfbox{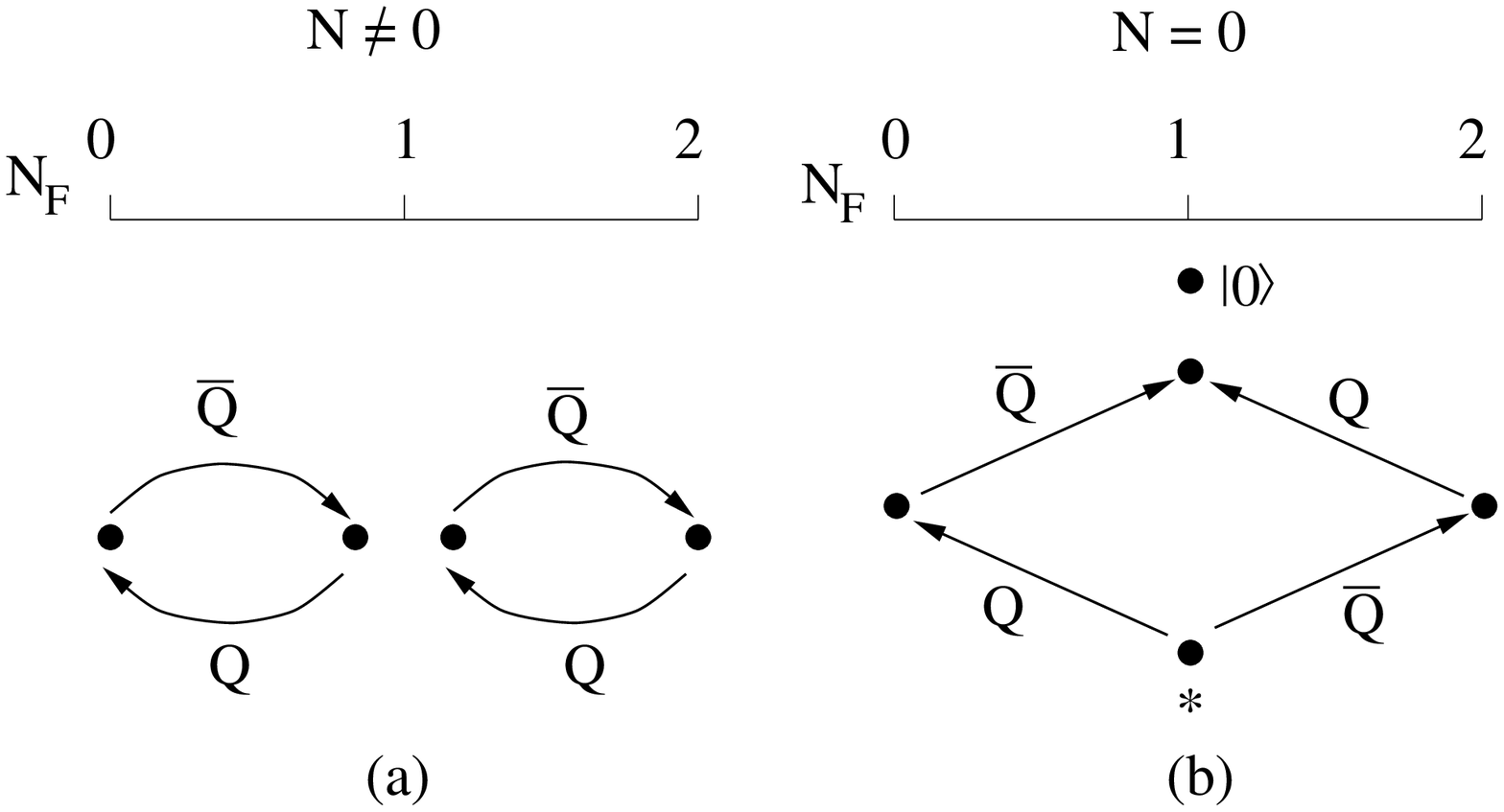} \vspace{15pt} {Fig.~1:
    Organization of multiplets in the SUSY Hamiltonian.  (a):
    Doublets, involving one bosonic and one Fermionic state for total
    number $N \ne 0$. (b) Singlet ground state $|0\rangle$ and
    quadruplets, with two Fermionic and two bosonic states for $N=0$.
    See text for details.}
\label{fig1}
\end{figure}
For $N\neq 0$, the states are in fact organized into doublets, as
indicated graphically in Fig.~1a. One class of doublets includes the
ladder of states with $N_F=0$,
\begin{eqnarray}
  \overline{Q}|N,0,0,N,E\rangle & = & c|N-1,1,N,0,E\rangle, \\
  Q|N-1,1,N,0,E\rangle & = & (N/c)|N,0,0,N,E\rangle, \\
  Q|N,0,0,N,E\rangle & = & \overline{Q}|N-1,1,N,0,E\rangle = 0, 
\end{eqnarray}
while another includes the states with $N_F=2$,
\begin{eqnarray}
  Q|N-2,2,N,0,E\rangle & = & c|N-1,1,0,N,E\rangle, \\
  \overline{Q}|N-1,1,0,N,E\rangle & = & (N/c)|N-2,2,N,0,E\rangle, \\
  \overline{Q}|N-2,2,N,0,E\rangle & = & Q|N-1,1,0,N,E\rangle = 0,
\end{eqnarray}
where different constants $c$ can appear in different expressions.

For $N=0$, things are more complicated.  Distinct eigenstates can be
found in the sectors with $N_F=-N_B = 0,2$, and for the ladder of
states with $N_F=-N_B=1$ and $\Gamma=\overline{\Gamma}=0$.  The fourth
subspace, spanned by the $|-1,1,*,*,n\rangle$ basis, however, {\sl is
  not closed under the action of $H$}.  The best that one can hope to
achieve in this sector, therefore, is to find an eigenstate of the
Hamiltonian projected back onto the same sector.  Such a state
$|-1,1,*,*,E\rangle$ satisfies
\begin{equation}
  H |-1,1,*,*,E\rangle = E |-1,1,*,*,E\rangle + |\psi\rangle,
  \label{pseudo_eigenstate}
\end{equation}
where 
\begin{equation}
  |\psi\rangle = \sum_n \tilde{\chi}_n |-1,1,0,0,n\rangle,
\end{equation}
and is hence orthogonal to $|-1,1,*,*,n\rangle$.  

Taken together with the other three states of energy $E$,
$|-1,1,*,*,E\rangle$ forms the fourth member of a super-quadruplet
(see Fig.~1b).  Using Eq.~(\ref{pseudo_eigenstate}),
it is straightforward to see that the three true eigenstates may be
obtained through the action of $Q$ and $\overline{Q}$ on
$|-1,1,*,*,E\rangle$.  In particular,
\begin{eqnarray}
  Q|-1,1,*,*,E\rangle & = & c_1 |0,0,0,0,E\rangle, \\
  \overline{Q}|-1,1,*,*,E\rangle & = & c_2 |-2,2,0,0,E\rangle, \\
  \overline{Q}|0,0,0,0,E\rangle & = & c_3 |-1,1,0,0,E\rangle, \\
  Q|-2,2,0,0,E\rangle & = & c_4|-1,1,0,0,E\rangle, \\
  \overline{Q}|-2,2,0,0,E\rangle & = & Q|0,0,0,0,E\rangle \nonumber \\
  =\overline{Q}|-1,1,0,0,E\rangle & = & Q|-1,1,0,0,E\rangle = 0.
\end{eqnarray}

We expect the exception to this rule to be a unique ground state with
zero energy, forming a super-singlet.  We can essentially pin down the
sector in which such a singlet can occur.  It cannot occur for $N\neq
0$, since $\Gamma + \overline{\Gamma} = N \neq 0$, and hence either
$Q$ or $\overline{Q}$ would create a new degenerate state.  For $N=0$,
it is possible only for a state in the $N_F = -N_B = 1$,
$\Gamma=\overline{\Gamma}=0$ ladder.  This is consistent with the
existence of one additional state $|-1,1,0,0,n=0\rangle$ in this
basis. 

\subsubsection{partition function and non-hermiticity}

Thus far, we have described the organization of eigenstates.  To
calculate physical quantities, we will need to perform a trace over the
Hilbert space of $e^{-LH}$.  Once the Hamiltonian is diagonalized,
this trace can be performed separately in each multiplet.  Some care
must be taken to account, however, for the non-Hermiticity of $H$.  

To do so, let us consider the behavior of the currents under hermitian
conjugation.  The usual SU(2) spin generators $\bbox{S}$ are of course
Hermitian.  The bosonic currents $\bbox{J}$ are, however, mixed.  In
particular $J^z$ is hermitian while $J^x$ and $J^y$ are
anti-Hermitian.  Since the Hamiltonian involves only ${\cal J}^z$ and
${\cal J}^x$, the net effect of conjugation is to change the sign of
$J^x$.  This can in fact be accomplished by the unitary transformation
($U = U^\dagger$)
\begin{equation}
  U = e^{i\pi J^z},
\label{Unitary} 
\end{equation}
which rotates
$J^x$
\begin{equation}
  U^\dagger J^x U = -J^x
\end{equation}
and therefore conjugates the Hamiltonian
\begin{equation}
  U^\dagger H U = H^\dagger.
\end{equation}
It therefore follows that if $|E\rangle_R$ is a right eigenstate of
$H$ with energy $E$, 
\begin{equation}
  |E\rangle_L = c U |E\rangle_R
\end{equation}
is a left-eigenstate of $H$, i.e.
\begin{equation}
  H^\dagger |E\rangle_L = E |E\rangle_L \;\; \Leftrightarrow\;\;
  \left._L\langle E|\right. H = \left._L\langle E|\right. E.
\end{equation}

Within each of the doublets, it is advantageous to normalize the
states such that
\begin{equation}
  \left._L\langle E|\right.E\rangle_R = 1.
\end{equation}
\end{multicols}
Left and right eigenstates with different quantum numbers are
automatically orthogonal, so this normalization guarantees the
resolutions of the identity
\begin{eqnarray}
  1_{N;N_F=0,1}  & = & \sum_{E} \bigg[ |N,0,0,N,E\rangle_R
  \left._L\langle N,0,0,N,E|\right. + |N-1,1,N,0,E\rangle_R
   \left._L\langle N-1,1,N,0,E|\right. 
   \bigg], \label{ri1} \\
  1_{N;N_F=1,2}  & = & \sum_{E} \bigg[ |N-2,2,N,0,E\rangle_R
  \left._L\langle N-2,2,N,0,E|\right. + |N-1,1,0,N,E\rangle_R
  \left._L\langle N-1,1,0,N,E|\right. \bigg], \label{ri2}
\end{eqnarray}
for $N\neq 0$.  Inserting these into the usual expression for the
trace in terms of an orthonormal basis gives
\begin{eqnarray}
  {\rm STr}_{N;N_F=0,1} O & = & \sum_{E} \bigg[ - \left._L\langle
    N,0,0,N,E|\right. O |N,0,0,N,E\rangle_R
   + \left._L\langle N-1,1,N,0,E|\right. O
  |N-1,1,N,0,E\rangle_R \bigg], \\
  {\rm STr}_{N;N_F=1,2} O & = & \sum_{E} \bigg[ -\left._L\langle
    N-2,2,N,0,E|\right. O |N-2,2,N,0,E\rangle_R
   + \left._L\langle N-1,1,0,N,E|\right. O
  |N-1,1,0,N,E\rangle_R \bigg],
\end{eqnarray}
\begin{multicols}{2}
\noindent where the minus signs arise from the $(-1)^{N_f}$ in the definition
of the supertrace, and again this holds only for $N\neq 0$.
Similarly, the ground state gives a unique contribution
\begin{equation}
  {\rm STr}_0 O = \left._L \langle 0|\right.O|0\rangle_R ,  
\end{equation}
where $|0\rangle \equiv |-1,1,0,0,E=0\rangle$ is the ground state.

The quadruplets are somewhat trickier.  The two states with $s=0$
(i.e. $N_F=0,2$) in each quadruplet can be separately orthogonalized,
i.e. 
\begin{eqnarray}
  \left._L\langle 0,0,0,0,E|\right. 0,0,0,0,E\rangle_R & = & 1, \\
  \left._L\langle -2,2,0,0,E|\right. -2,2,0,0,E\rangle_R & = & 1,
\end{eqnarray}
since $U|-2,2,0,0,E\rangle_R$ remains in the subspace with $N_F=-N_B =
2$, and similarly for the $N_F=N_B=0$ state.  The same is not true for
the two states with $N_F=-N_B=1$.  In fact, it is straightforward to
show that corresponding left and right eigenstates are actually
orthogonal!  That is,
\begin{eqnarray}
   \left._L\langle -1,1,0,0,E|\right. -1,1,0,0,E\rangle_R & = & 0,  \\
   \left._L\langle -1,1,*,*,E|\right. -1,1,*,*,E\rangle_R & = & 0,
\end{eqnarray}
for $E\neq 0$.  This is because the $U$ operator takes each basis
state from one subspace into the other, as can be seen directly from
Eqs.~(\ref{zero_ladder},\ref{star_ladder}).  Instead of the usual
normalization condition, therefore, we must require
\begin{eqnarray}
  \left._L\langle -1,1,0,0,E|\right. -1,1,*,*,E\rangle_R & = & 1,  \\
   \left._L\langle -1,1,*,*,E|\right. -1,1,0,0,E\rangle_R & = & 1.
\end{eqnarray}
\end{multicols}
The corresponding resolution of the identity is
\begin{eqnarray}
  1_{N=0} & = & \sum_E \bigg[ |2,-2,0,0,E\rangle_R
  \left._L\langle 2,0,0,0,E|\right. 
  + |0,0,0,0,E\rangle_R \left._L\langle 0,0,0,0,E|\right. \nonumber \\
  & & + |-1,1,0,0,E\rangle_R \left._L \langle -1,1,*,*,E| \right.
    + |-1,1,*,*,E\rangle_R \left._L \langle -1,1,0,0,E| \right. \bigg].
\end{eqnarray}
$ $From this identity follows the expression for the trace,
\begin{eqnarray}
  {\rm STr}_{N=0} O & = & \sum_E \bigg[ -\left._L \langle
    2,-2,0,0,E|\right. O |2,-2,0,0,E\rangle_R 
  - \left._L \langle
    0,0,0,0,E|\right. O |0,0,0,0,E\rangle_R \nonumber \\
  & & + \left._L \langle
    -1,1,0,0,E|\right. O |-1,1,*,*,E\rangle_R
  + \left._L \langle
    -1,1,*,*,E|\right. O |-1,1,0,0,E\rangle_R \bigg]  .
\end{eqnarray}
\begin{multicols}{2}

In this paper we are studying primarily properties of the system in
the thermodynamic limit $L\rightarrow\infty$.  In this case the
(super)trace of $e^{-LH}$ reduces to an expectation value in the
singlet ground state, i.e.
\begin{equation}
  \langle O \rangle = {\rm STr} O e^{-LH} \rightarrow \left._L \langle
    0|\right. O|0\rangle_R . 
\end{equation}

It is worth noting, however, that SUSY indeed implies that the
partition function itself is exactly one, even in the finite system.
This is because each multiplet other than the singlet contains equal
numbers of fermionic ($N_f$ odd) and bosonic ($N_f$ even) states,
which thereby cancel in the supertrace.  All that remains is the
expectation value in the zero-energy ground state, which is unity even
for $L$ finite.

\section{Ground State Properties}

\subsection{Schr\"odinger Equation}

As we saw in the previous section, the ground state resides in a
unique (non-unitary) singlet representation of SU(2) algebra of the
total current $\bbox{\cal J}$.  As such, it must be annihilated by
both $Q$ and $\overline{Q}$, and hence fits into the subspace with
$N_F=-N_B=1$, $\Gamma=\overline{\Gamma}=0$, i.e.
\begin{equation}
  |0\rangle_R = \sum_{n=0}^\infty \phi_n |-1,1,0,0,n\rangle.
\end{equation}
In this basis, ${\cal J}^x$ and ${\cal J}^z$ have the very simple
matrix elements
\begin{eqnarray}
  {\cal J}^z |n\rangle_0 & = & n|n\rangle_0, \\
  {\cal J}^x |n\rangle_0 & = & {1 \over 2} \Bigg[
  (n+1)|n+1\rangle_0 - (n-1)|n-1\rangle_0 \Bigg],
\end{eqnarray}
where we have introduced the more compact notation $|n\rangle_0 \equiv
|-1,1,0,0,n\rangle$.  Using these results, and the requirement of a
zero energy ground state,
\begin{equation}
  H|0\rangle_R = E|0\rangle_R = 0,
\end{equation}
we obtain straightforwardly the Schr\"odinger equation
\end{multicols}
\begin{equation}
n\Bigg[ -(n+1)\phi_{n+2} + 2n \phi_n - (n-1)\phi_{n-2} - m_0 \left(
  \phi_{n+1} - \phi_{n-1}\right) + 2\omega \phi_n\Bigg] = 0.
\end{equation}
Which implies
\begin{equation}
  -(n+1)\phi_{n+2} + 2n \phi_n - (n-1)\phi_{n-2} - 2M \left(
  \phi_{n+1} - \phi_{n-1}\right) + 2\omega \phi_n = 0, \;\;\; {\rm
  for} \; n \neq 0.
\label{SchrodE0}
\end{equation}
Here we have defined $M = m_0/2$ to simplify some expressions in what
follows. 

\begin{multicols}{2}

\subsection{Normalization and density of states}

$ $From the previous section, given the right eigenstate $|0\rangle_R$, a
corresponding left eigenstate is obtained by
\begin{equation}
  |0\rangle_L = c U |0\rangle_R = (-1)^{J^z-1/2} |0\rangle_R,
\end{equation}
with an appropriate choice for the constant $c$.  Applying this to the
basis of ${\cal J}^z$ eigenstates gives
\begin{equation}
  (-1)^{J^z-1/2}|n\rangle_0 = \cases{ (-1)^n |-1,1,*,*,n\rangle &
    $n>0$ \cr |0\rangle_0 & $n=0$ \cr}.
\end{equation}
Since the $|n\rangle_0$ and $|-1,1,*,*,n\rangle$ states are
orthogonal, the normalization requirement reduces to
\begin{equation}
  \left._L \langle 0|\right. 0\rangle_R = |\phi_0|^2 = 1.
\label{norm_cond}
\end{equation}

The density of states is obtained in a similar way.  
From Eq.~(\ref{DOS}) the density of states
is obtained by analytically continuing,
\begin{equation}
 \overline{{\cal G}}(x=0;i\omega) =  \sum_\alpha \overline{G}_{\alpha\alpha}(0,i\omega) = -2 i \left._L
    \langle 0| \right. S^z |0\rangle_R ,
\end{equation}
where the second equality follows from Eq.~(\ref{GSTr}).
Simple manipulations then give
\begin{equation}
 \overline{{\cal G}}(0;i\omega) =
  i \sum_{n=0}^\infty (-1)^n |\phi_n|^2. 
  \label{dos}
\end{equation}

\subsection{Continuum limit and boundary conditions}

In order to solve for the ground state wavefunction, first note the
following simple property:  if $\phi^+_n(M)$ is a solution of
Eq.~(\ref{SchrodE0}), then so is 
\begin{equation}
  \phi_n^-(M) = (-1)^n \phi^+_n(-M).
\end{equation}
Since this is a fourth-order linear difference equation, we would
expect four linearly independent solutions.  The above result reduces
these to two trivially related pairs.  To understand the nature of
these two solutions, it is instructive to consider the limit
$M=\omega=0$.  In this case there are two obvious solutions: $\phi_n
= 1$ and $\phi_n = (-1)^n$ satisfy the $E=0$ Schr\"odinger equation
(\ref{SchrodE0}).  However, these wavefunction are {\it not}
normalizable, a feature due to a pathology of the theory {\it at}
$\omega = 0$.  Indeed, non-zero $\omega$ is essential in guaranteeing
convergence of the bosonic functional integral in the generating
function.  Moreover, we are interested in behavior off criticality
with $M \ne 0$ and at finite energy $i \omega \rightarrow \epsilon$.

A general solution of the Schr\"odinger equation Eq.~(\ref{SchrodE0})
is daunting (although possible for $\omega \ne 0$  - 
see Appendix B), so for simplicity we focus our attention on the
critical regime very close to the localization transition
where both 
$\omega, M \ll g = 1$.  
In this scaling regime one expects the wavefunction
to remain close to a superposition of the two trivial (constant and
$(-1)^n$) solutions, i.e.
\begin{equation}
  \phi_n = c_1\phi(n,M) + c_2 (-1)^n \phi(n,-M),
  \label{superposition}
\end{equation}
where $\phi(n,M)$ is slowly varying with $|\phi(n+1,M) - \phi(n,M)| \ll
|\phi(n,M)|$.   This suggests a ``continuum" limit,
in which  $\phi(n,M)$ may be regarded as a continuous function of $n$,
and discrete differences in the Schr\"odinger equation
are replaced with derivatives.
In this continuum approximation the Schr\"odinger equation Eq.~(\ref{SchrodE0})
becomes,
\begin{equation}
n {{d^2\phi} \over {dn^2}} + (1+M){{d\phi} \over {dn}} = {\omega \over
  2}\phi.
\label{continuum_eqn}
\end{equation}

For $\omega, M \ll 1$ the solution of this continuum equation is
expected to coincide with the exact eigenfunction of Eq.~(\ref{SchrodE0})
for $n\gg 1$.  In Appendix B, we verify this for the special case
$M=0$, where it is possible to solve directly the discrete
Schr\"odinger equation.

The continuum differential equation (\ref{continuum_eqn}) must be
supplemented by an appropriate boundary condition.  A natural physical
requirement is that $\phi(n,M) \rightarrow 0$ for $n \rightarrow
\infty$.  Because we have still have the freedom to choose $c_1$ and
$c_2$ in Eq.~(\ref{superposition}), the normalization at the origin
can remain arbitrary at this stage.

The two constants in Eq.~(\ref{superposition}) then require two
additional constraints.  The first comes from Eq.~(\ref{norm_cond}),
$\phi_0=1$, giving
\begin{equation}
  c_1 \phi(1,M) + c_2 \phi(1,-M) = 1.
  \label{constraint1}
\end{equation}
The second constraint is obtained from
Eq.~(\ref{SchrodE0}) for $n=1$:
\begin{equation}
  \phi_3 - \phi_1  =  -M(\phi_2-\phi_0) + \omega\phi_1
\end{equation}
In the limit $M,\omega \ll 1$,
and using Eq.~(\ref{superposition}), this becomes
\begin{equation}
  c_1 \phi'(1,M) = c_2 \phi'(1,-M) .
  \label{constraint2}
\end{equation}

\subsection{Solution}

We are now in a position to obtain the solution to the continuum
equation.
Under an exponential change of variables,
$n = e^z$, with $\Phi(z,M) = \phi(e^z,M)$,
Eq.~(\ref{continuum_eqn})  takes a more illuminating form,
\begin{equation}
\left[ -{{d^2} \over {dz^2}} - M {d \over {dz}} + {\omega \over 2} e^z
\right]\Phi_\pm = 0.
\label{Liouville}
\end{equation}
Indeed, for $M=0$ this is equivalent to a Schr\"odinger equation
for a particle moving in an exponential potential.
Since $z=\ln n$, the domain of the equation is $0\leq z
<\infty$.  For small $\omega$, however, the potential is negligible
for small $z$.  It rises very abruptly and becomes of order one for $z
\sim z_w$, where
\begin{equation}
z_w = |\ln\omega|.
\end{equation} 
In the small $\omega$ limit of interest, then, there is a region of
divergent width over which Eq.~(\ref{Liouville}) effectively
describes a {\it free} particle.  The width of the potential is,
however, only of order one.  Thus we expect that on the scale of
$z_w$, the exponential potential can be well approximated by a {\it
  hard wall}.

In the following we make this hard-wall approximation,
replacing the continuum equation (\ref{Liouville}) by
\begin{equation}
  \left[ -{{d^2} \over {dz^2}} - {M} {d \over {dz}} \right]\Phi = 0,
  \label{Modified_diffusion}
\end{equation}
with $0 \le z \le z_w$. and the boundary condition
\begin{equation}
  \Phi(z_w) = 0. 
\label{bc}
\end{equation}
This simple equation can be readily solved (below) and
the density of states extracted.

As shown in Appendix C, the hard-wall approximation is not necessary,
since the exact continuum equations can be solved explicitly.
Although the hard-wall wave function differs from the exact solution,
the resulting density of states coincides, up to an overall
(non-universal) multiplicative constant.

The solution of Eq.~(\ref{Modified_diffusion}) consistent with 
the hard-wall boundary condition (Eq.~(\ref{bc})) is
\begin{equation}
  \Phi(z,M) ={1 \over {Mz_w}}\left[e^{-M z} - e^{-M z_w}\right],
  \label{Phi_solution}
\end{equation}
where we have assumed that $\omega,{M} \ll 1$, and chosen a convenient
(but arbitrary) normalization.  As $M \rightarrow 0$ this reduces to
\begin{equation}
  \Phi(z) = 1 - {z \over z_w}.
\end{equation}
Applying the constraints in
Eqs.~(\ref{constraint1},\ref{constraint2}), determines
the constants as
\begin{equation}
  c_1 = c_2 = {{z_w M\omega^M} \over {1-\omega^{2M}}}.
\end{equation}

To extract the density of states, we use Eq.~(\ref{dos}) to write,
\begin{eqnarray}
  \overline{{\cal G}}(0,i\omega)  & = & 2i c_1 c_2
   \int_1^\infty \! dn \,
   \phi(n,M)\phi(n,-M) \nonumber \\
   & & = 2 i c_1 c_2\int_0^{z_w}\!
  dz \, e^z \Phi(z,M) \Phi(z,-M).
  \label{intDOS}
\end{eqnarray}
The integral can be readily performed
giving,
\begin{equation}
  \overline{{\cal G}}(0,i\omega)  = i {{{M}^2} \over \omega}
  f(\omega^{M}), 
  \label{exactG}
\end{equation}
with the {\it exact} scaling function
\begin{equation}
  f(Y) = c\left({Y \over {1-Y^2}}\right)^2 .
  \label{hard_wall_sf}
\end{equation}
Our present implementation of the hard-wall approximation reproduces
the exact value of the non-universal constant, $c=4$, obtained in Appendix
C, but this result depends upon the precise position of the wall.

An exact expression for the density of states in the critical regime
can now be obtained from
Eq.~(\ref{DOS}) by performing an analytic continuation,
$\rho(\epsilon) = (1/\pi) Im \overline{{\cal G}}(0;\epsilon + i0^+)$.
Noting that $\overline{{\cal G}}(0,i\omega)$ is pure imaginary and odd in $\omega$,
one readily obtains (for $\epsilon >0$),
\begin{equation}
  \rho(\epsilon) = {{M}^3 \over \epsilon} g(\epsilon^{M}),
\end{equation}
with
\begin{equation}
  g(Y) = {1\over 2}Y f'(Y) = c {{Y^2(1+Y^2)} \over {(1-Y^2)^3}}.
\end{equation}
In the $M \rightarrow 0$ limit, this reduces to,
\begin{equation}
  \rho(\epsilon) \sim {1 \over {\epsilon|\ln\epsilon|^3}},
\end{equation}
a result obtained previously by other methods.

A special feature of 1d, is that the {\it typical} localization length,
$\tilde{\xi}$, can be extracted from the real part of the Green's
function {\it at} $x=0$.  As derived in Ref.~\onlinecite{Thouless72},
$\tilde{\xi}(\epsilon)$ satisfies
\begin{equation}
{ \partial \tilde{\xi}^{-1}  \over  \partial \epsilon } = 
P \int d\epsilon^\prime {\rho(\epsilon^\prime)  \over  
\epsilon-\epsilon^\prime }  = -{1 \over \pi} Re \overline{{\cal G}}(x=0;\epsilon)   .
\end{equation}
Performing an analytic continuation to real energy
using Eq.~(\ref{exactG}) gives $Re \overline{{\cal G}}(0;\epsilon)
\sim - 1/\epsilon|\ln\epsilon|^2$.  Integration from
$\epsilon =0$ using the fact that
$\tilde{\xi}^{-1}(0) =0$ gives
the result $\tilde{\xi} \sim |\ln\epsilon|$.

\section{Fermion Green's function}

To determine the mean correlation length, we need to calculate
the Green's function, $\overline{G}(x,i\omega)$ at {\it non-zero} $x$.
From Eq.~(\ref{GSTr}) this takes the form
\begin{equation}
  \overline{G}_{\alpha\beta}(x,i\omega) = i (-1)^\alpha \left._L \langle 0
    |\right. F_\alpha e^{-x H} F_\beta^\dagger |0\rangle_R.
\end{equation}
Given the quantum numbers of the ground state, the state
$F_\beta^\dagger |0\rangle_R$ clearly has $N_B=-1$, $N_F=2$, $\Gamma =
N = 1$, $\overline{\Gamma}=0$.  We can therefore insert the resolution
of the identity from Eq.~(\ref{ri2}) to give
\end{multicols}
\begin{equation}
  \overline{G}_{\alpha\beta}(x,i\omega) = i (-1)^\alpha \sum_E \left._L
    \langle 0 |\right. F_\alpha |-1,2,1,0,E\rangle_R \left._L \langle
  \right. -1,2,1,0,E|F_\beta^\dagger |0\rangle_R e^{- E x}. 
  \label{GF_expansion}
\end{equation}
Our task is thus to determine the matrix elements and eigenvalues
needed to carry out this sum.  To do so, we expand the eigenstate in
the appropriate basis,
\begin{equation}
  |-1,2,1,0,E\rangle_R = \sum_{n=0}^\infty \chi_n |n\rangle_1,
\end{equation}
where we have abbreviated 
\begin{equation}
  |n\rangle_1 \equiv |-1,2,1,0,1/2+n\rangle.
\end{equation}
The wavefunction $\chi_n$ then satisfies the
Schr\"odinger wave equation,

\begin{eqnarray}
  2\omega\left[n+{1 \over 2}\right] \chi_n && + 2M
  \left[n\chi_{n-1} - (n+1)\chi_{n+1}\right] \nonumber \\
  && - (n+2)(n+1)\chi_{n+2} + (2n^2+2n+1)\chi_n-
  n(n-1)\chi_{n-2} = E\chi_n,
\label{spin_zero_discrete_SE}
\end{eqnarray}

\begin{multicols}{2}
\noindent where we have again set $g=1$.  Again, this equation has the
property that multiplication of a solution by $(-1)^n$ yields a
solution for $M \rightarrow -M$.  We therefore expect
\begin{equation}
  \chi_n(M,E) = c_3 \chi(n,M,E) + c_4 (-1)^n \chi(n,-M,E).
  \label{chi_decomposition}
\end{equation}

Naively, $\chi(n,M,E)$ can be obtained in the continuum limit.
Converting finite differences to derivatives, we obtain
\end{multicols}
\begin{equation}
  \left[ - \left(2n {d \over {dn}} +1\right)^2 - 2M\left(2n{d \over
        {dn}}+1\right) 
    + 2\omega n \right] \chi(n,M,E)= E\chi(n,M,E).
  \label{continuum_singlet}
\end{equation}
Upon tranforming to logarithmic variables,
\begin{equation}
  \chi(n,M) = \Phi_E (z = \ln n,M),
\end{equation}
one has
\begin{equation}
  \left[ - \left(2\partial_z +1\right)^2 - 2{M}\left(2\partial_z+1\right)
    + 2\omega e^z \right]\Phi_E(z,M) = E\Phi_E(z,M).
\label{excited_continuum}
\end{equation}
\begin{multicols}{2}

Based on previous experience, we expect that the
hard-wall approximation gives exact results in the scaling limit.
In Appendix D we verify this explicitly, by constructing
exact solutions of the continuum equation (\ref{continuum_singlet}).
Within the hard-wall approximation
the potential $2\omega e^z \rightarrow 0$ in
Eq.~(\ref{excited_continuum}) is dropped,
and replaced by a boundary condition at
$z_w = |\ln\omega|$,
\begin{equation}
  \Phi_E(z_w,M) = 0.
\end{equation}

The general solution of Eq.~(\ref{excited_continuum}) with the
hard-wall boundary condition is
\begin{equation}
  \Phi_E(z) = e^{-(1+M)z/2} \sin \left( \beta (z-z_w)/2 \right),
  \label{standing_wave}
\end{equation}
where $\beta = \sqrt{E-M^2}$.  

So far, we have not determined the spectrum , or allowed values of
$\beta$.  In an ordinary quantum problem, these eigenvalues would be
fixed by a boundary condition at $z=0$.  In this case, however, such a
simple treatment is problematic.  The difficulty arises because,
unlike in the ground state sector, neither $\chi_n = 1$ nor $\chi_n =
(-1)^n$ are solutions in the limit $\omega = M = 0$.  We therefore
expect a non-trivial solution for $n = O(1)$, in which the
discreteness of $n$ is important and the continuum limit is {\sl not}
valid.

Fortunately, for $1 \lesssim n \ll 1/\omega$, we can 
obtain an asymptotic approximation which does not rely upon the
continuum limit.  This is possible because for $n \ll 1/\omega$, the
$\omega {\cal J}^z$ term in the Hamiltonian Eq.~(\ref{Hspin})
can be regarded as a small
perturbation.  Neglecting this term, $H$ is a function only of ${\cal
  J}^x$, so that the eigenfunctions of $H$ are simply eigenfunctions
of ${\cal J}^x$.  As a first step, consider the state $|\alpha\rangle$
with
\begin{equation}
  {\cal J}^x |\alpha\rangle = i\alpha |\alpha\rangle.
\end{equation}
Expanding $|\alpha\rangle$ in the ${\cal J}^z$ basis,
\begin{equation}
  |\alpha\rangle = \sum_n \psi_n(\alpha)|n\rangle_1,
\end{equation}
one finds the simpler Schr\"odinger equation
\begin{equation}
  (n+1)\psi_{n+1}(\alpha) - n \psi_{n-1}(\alpha) = -2i\alpha
  \psi_n(\alpha).
  \label{Jx_eigenstates}
\end{equation}
This equation can be solved exactly (see Appendix E).  For large $n$,
the solution which is well-behaved at the origin behaves asyptotically
as
\end{multicols}
\begin{equation}
  \psi_n(\alpha) \sim n^{-1/2} \left[
    (2n)^{-i\alpha} \Gamma(1/2+i\alpha) + (-1)^n (2n)^{i\alpha}
    \Gamma(1/2-i\alpha) \right] . \label{Jx_asymptotic} 
\end{equation}

\begin{multicols}{2}
Eigenstates of $H$ therefore take the form
\begin{equation}
  \chi_n \sim c_+ \psi_n(\alpha_+) + c_- \psi_n(\alpha_-),
  \;\;\; {\rm for}\; 1 \ll n \ll 1/\omega,
  \label{outer}
\end{equation}
where $\alpha_\pm$ are the two roots of the equation $4\alpha^2 +
4iM\alpha = E$, i.e.
\begin{equation}
  \alpha_\pm = {{-i M \pm \beta} \over 2},
\end{equation}
and $\beta = \sqrt{E-M^2}$.

Mathematically, Eq.~(\ref{outer}) is an {\sl outer} solution, valid
outside a boundary layer which occurs for large $n$.  To obtain a
complete solution, it must be matched to the {\sl inner} solution,
valid ``inside'' the boundary layer, which is just the continuum
regime of large $n$.  Within
the hard-wall approximation, this is just the standing-wave in
Eq.~(\ref{standing_wave}).

To match the two solutions, we let $n=e^z$ in
Eqs.~(\ref{Jx_asymptotic},\ref{outer}), which gives
\end{multicols}
\begin{equation}
  \chi_n \sim e^{-z/2} \left\{ e^{-M z/2} \left[ c_+ e^{-i\beta z/2} +
      c_- e^{i\beta z/2}\right] + (-1)^n e^{M z/2} \left[ c_+ e^{i\beta z/2} +
      c_- e^{-i\beta z/2}\right] \right\}.
\end{equation}
Similarly, using Eqs.~(\ref{chi_decomposition},\ref{standing_wave}),
the continuum solution gives 
\begin{equation}
  \chi_n \sim e^{-z/2}\left\{ c_3 e^{-M z/2} + (-1)^n c_4 e^{M
      z/2}\right\} \sin \left( \beta (z-z_w)/2 \right).
\end{equation}
\begin{multicols}{2}
These expressions are equal in two situations.  One can take $c_+ =
c_- = c_3/2= c_4/2$ if $\beta z_w$ is an odd multiple of $\pi$.
Alternatively,  $c_+ = -c_- = c_3/2 = -c_4/2$ if $\beta z_w$ is
an even multiple of $\pi$.  The final, matched solution for both cases
can thus be written
\begin{equation}
  \Phi^{(k)}(z,M) \equiv \Phi_{E_k}(z,M) = e^{-(1+M)z/2} \sin \left(
    \beta_k z/2 + \theta_k\right),
\end{equation}
where 
\begin{eqnarray}
  \beta_k & = & \pi k/z_w, \\ 
  E_k & = & M^2 + \left( {{\pi k} \over z_w} \right)^2, \\
  \theta_k & = & \cases{0 & $k$ even \cr \pi/2 & $k$ odd \cr}, \\
  c_3 & = & (-1)^{k+1} c_4 \equiv c.
\end{eqnarray}
The discrete quantum number $k=1,2,...\infty$.

The corresponding left eigenstate is obtain by acting with $U$
from Eq.~(\ref{Unitary}).
Defining
\begin{equation}
|k \rangle_{R/L} = |-1,2,1,0,E_k\rangle_{R/L}  ,
\end{equation}
we choose
\begin{equation}
  |k\rangle_L = (-1)^{k+1} e^{i\pi (J^z-1/2)}|k\rangle_R.
\end{equation}
With this choice, normalization implies that
the constant
$c=z_w^{-1/2}$.  Thus the final expressions for the 
(normalized) eigenstates are
\end{multicols}
\begin{eqnarray}
  |k\rangle_R & = &  z_w^{-1/2}\sum_n
  \left[ \chi(n,M) + (-1)^{n+k+1} \chi(n,-M)\right] |n\rangle_1, \\
  |k\rangle_L & = & z_w^{-1/2}\sum_n
  \left[ (-1)^{n+k+1} \chi(n,M) + \chi(n,-M)\right] |n\rangle_1. 
\end{eqnarray}
\begin{multicols}{2}
These actually form an orthonormal set
\begin{equation}
  \left._L \langle \right. k|k'\rangle_R = \delta_{kk'},
\end{equation}
as can be verified by direct computation.

\end{multicols}
Having obtained the full set of eigenvalues and left and right
eigenfunctions of $H$ in the appropriate sector, it is a simple matter
to evaluate the Green's function using Eq.~(\ref{GF_expansion}).
Using the hard-wall eigenfunctions, we find
\begin{eqnarray}
  \overline{G}_{\alpha\beta}(x,i\omega) & = & i {32\pi^2 \over {\omega
      |\ln\omega|^3}} {M^2\omega^{2M} \over (1-\omega^{2M})^2} 
  (-1)^\alpha \nonumber \\
  & & \times \sum_{k=1}^\infty (-1)^{k+1}k^2 \left[(-1)^\alpha
    \omega^{M/2} + (-1)^{k+1} \omega^{-M/2}\right] \left[ (-1)^\beta
    \omega^{M/2} + (-1)^{k+1} \omega^{-M/2}\right] e^{-\pi^2 k^2
    x/|\ln\omega|^2}.
\end{eqnarray}
\begin{multicols}{2}
One thereby obtains the final form for the Green's function,
exact in the scaling limit:
\begin{equation}
  \overline{\cal G}(x,i\omega) = \cases{ {i^{x+1} \over
      {\omega|\ln\omega|^5}} f^e_\omega(\omega^{M})
    F^e_\omega(x/\xi_\omega) e^{-x/\xi_{M}}, & $x$ even \cr
    {i^{x+1} \over
      {\omega|\ln\omega|^5}} f^o_\omega(\omega^{M})
    F^o_\omega(x/\xi_\omega) e^{-x/\xi_{M}}, & $x$ odd \cr},    
\label{G_omega_scaling}
\end{equation}
where the universal scaling functions are given by,
\begin{eqnarray}
  f^e_\omega(Y) & = & Y f^o_\omega(Y) = \left({{Y\ln Y} \over
      {1-Y^2}}\right)^2, \\ 
  F^e_\omega(Y) & = & A \sum_{k=1}^\infty k^2 e^{-k^2 Y}, \\
  F^o_\omega(Y) & = & A \sum_{k=1}^\infty (-1)^{k+1} k^2 e^{-k^2 Y}. 
  \label{f_scaling_fns}
\end{eqnarray}
Here the non-universal amplitude $A = 32\pi^2$ within the hard-wall
approximation, and we have defined two correlation lengths,
\begin{equation}
  \xi_{M}  = 1/{M}^2,  \qquad \xi_\omega = \left({{\ln\omega}
      \over \pi}\right)^2. \label{length_scales}
\end{equation}
Notice that $\overline{{\cal G}}(x;i\omega)$ is pure imaginary
for $x$ even and pure real for $x$ odd,
as dictated by particle-hole symmetry.  Eq.~(\ref{G_omega_scaling})
is valid for $\omega >0$.  Particle-hole symmetry
then determines the Green's function for $\omega < 0$, since
$\overline{{\cal G}}(x;i\omega)$
is odd in $\omega$ for even $x$ and even in $\omega$ for odd $x$.

Eq.~(\ref{length_scales}) gives us the correlation length exponent,
$\nu =2$, defined by $\xi \sim M^{-\nu}$, and the dynamical exponent,
$z=\infty$, defined by $\omega \sim \xi^z$.
Eq.~(\ref{G_omega_scaling}) is actually simpler than one would
generally expect on the basis of scaling -- the most general scaling
form would not factorize as it does here.

The correlator $C(\epsilon) = (1/\pi) {\rm Im} \overline{{\cal G}}(\epsilon + i0^+)$ can be obtained via analytic continuation, using the
symmetry properties of the Green's function under $\omega \rightarrow
- \omega$.  One finds
\begin{eqnarray}
  C(x=2n,\epsilon,M) & = & {{(-1)^n} \over {\epsilon|\ln\epsilon|^6}}
  F^e_\epsilon\left({x \over {\xi_\epsilon}},\epsilon^M\right)
  e^{-x/\xi_M}, \label{scaling1}\\
  C(x=2n+1,\epsilon,M) & = & {{(-1)^{n+1}} \over
    {\pi \epsilon|\ln\epsilon|^5}} F^o_\epsilon\left({x \over
      {\xi_\epsilon}},\epsilon^M\right) e^{-x/\xi_M},
\end{eqnarray}
where the energy dependent ``localization" length is
\begin{equation}
  \xi_\epsilon  =  \left({{\ln\epsilon} \over \pi}\right)^2 .
\end{equation}

The even sublattice scaling function no longer factors,
\begin{eqnarray}
  F^e_\epsilon(Y,Z) & = & f^e_\omega(Z)\left[{5 \over 2}F^e_\omega(Y) + Y
    F^{e\prime}_\omega(Y)\right] \nonumber \\
    & & + {1\over 2}Z\ln Z f^{e\prime}_\omega(Z) F^e_\omega(Y),
  \label{Fe_scalingfn}
\end{eqnarray}
while the odd sublattice scaling function remains simple,
\begin{equation}
  F^o_\epsilon(Y,Z) = F_\omega^o(Y) f_\omega^o(Z).
  \label{last_scalingfn}
\end{equation}

The scaling forms, Eqs.~(\ref{scaling1}--\ref{last_scalingfn}), encode
several significant physical properties.  First consider the same
sublattice correlation ($x=2n$), for simplicity at zero staggering
($M=0$).  For distances shorter than the correlation length, this has
a slow power-law decay, since $F_\epsilon^e(Y,1) \sim Y^{-3/2}$, for
$Y \ll 1$.  In particular,
\begin{equation}
  C(x=2n,\epsilon,M=0) \sim {{(-1)^n} \over {\epsilon|\ln\epsilon|^3}}
  {1 \over |x|^{3/2}},
\end{equation}
for $|x| \ll \xi_\epsilon$.  This can be understood as the product of
the density of states and a two-point ``wavefunction correlation'',
with multifractal scaling exponent (see the next section) $y(q=1) =
3/2$.  For distances $x \gg \xi_\epsilon$ (and $M=0$), even the rare
wavefunctions are localized, and $C(x)$ decays exponentially.  The
full scaling function, which describes the crossover between these two
limits, is plotted in Fig.~2.  Note the {\sl change of sign} for $Y
\approx 2.5$ -- this may be interpreted physically as arising from
the first node in the dominant rare wavefunction at energy $\epsilon$.
\begin{figure}[hbt]
  \setlength{\unitlength}{1.0in}
  \begin{picture}(4.0,4.0)(0.4,0)

    \put(0.5,0.5){
      \epsfxsize=3.0in\epsfbox{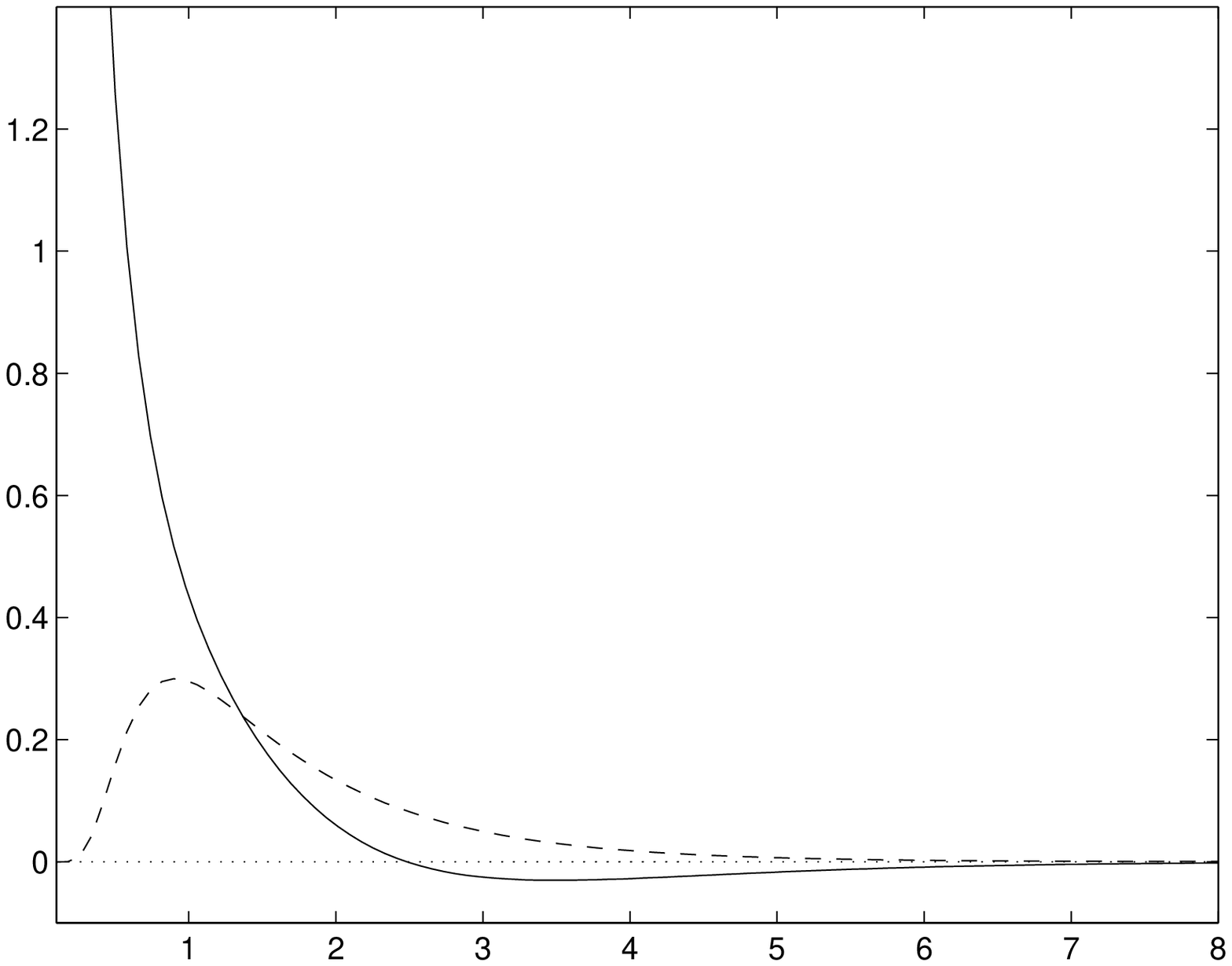}}
    \put(2.2,0.4){$Y$}
    \put(1.0,2.1){$F_\epsilon^{e}(Y)$}
    \put(1.5,1.0){$F_\epsilon^{o}(Y)$}
  \end{picture}
    {Fig.~2: Scaling functions for the Fermion Green's function.  The
      solid and dashed lines are the scaling functions,
      $F^e_\epsilon(Y)$ and $F^o_\epsilon(Y)$, for correlators $C(x)$
      between two sites on the same and different sublattices,
      respectively.}
\label{fig2}
\end{figure}
The different sublattice Green's function behaves quite differently at
short distances.  Using the Poisson summation formula, it is
straightforward to show that
\begin{equation}
  F_\epsilon^o(Y,1) \sim {\sqrt{\pi} \over {2 Y^{3/2}}} \left[1 -
    {\pi^2 \over {4Y}}\right]
    e^{-\pi^2/4Y} + O(e^{-9\pi^2/4Y}),
\end{equation}
for $Y \ll 1$.  This implies that $C(x=2n+1)$ is {\sl much smaller}
than the same sublattice correlator .  The suppression is due to the
fact that sublattice mixing of wavefunctions occurs only at finite
$\epsilon$.  The exact zero energy ``wavefunction'', discussed in
detail in the next section, lies {\sl entirely} on a single sublattice.

\section{Multifractal wavefunctions}
\label{sec:multifractal}

The above results reveal that the ensemble averaged single particle
Green's function decays as a power law with an exponent $3/2$ for
distances smaller than the correlation lenth.  When $m_0=0$, the
relevant scale is $\xi_\epsilon = |\ln\epsilon|^2/\pi^2$, which
diverges at the band center.  Such power law spatial scaling is in
striking contrast to the behavior of the {\it typical} Green's
function, which decays as a stretched exponential, $G_{\rm typ.} \sim
\exp(-ax^{1/2})$ at criticality.  This stretched exponential form
follows directly from the exact zero energy wave function,
Eq.~(\ref{wavefunction}), whose logarithm undergoes a one-dimensional
random walk.

Power law scaling of the {\it average} Green's function at criticality
can also be understood (albeit more subtly) in terms of
Eq.~(\ref{wavefunction}).  To see this it is useful to consider
ensemble averaged correlation functions of the zero energy
wavefunction.  As with higher dimensional localization transitions,
such as the plateau transition in the IQHE, this critical wavefunction
is expected to exhibit multi-fractal scaling characteristics.  As
shown in a recent paper, such correlators in this 1d case can be
computed exactly, via a mapping to Liouville quantum mechanics.  This
mapping exploits the equivalence between imaginary time quantum
mechanics and the one-dimensional random walk.  Slightly generalizing
this work, we compute below the full multi-fractal spectrum for the 1d
critical case.  This calculation is instructive since it reveals a
link between the supersymmetry calculations and Liouville quantum
mechanics.

To extract average wavefunction correlators, it is necessary to
consider normalized states.  We thus consider a finite system of
length $L$, and normalize the wavefunction from
Eq.~(\ref{wavefunction}) over the finite interval $|x| < L/2$.
Focussing on one component of the spinor wavefunction, say
$\phi_+(x)$, an appropriately normalized wavefunction, denoted as
$\psi(x)$, can be written,
\begin{equation}
\psi(x) = {\cal N}^{-1/2} e^{z(x)}  ,
\end{equation}
with normalization
\begin{equation}
{\cal N} = \int_{-L/2}^{L/2} \! dx\, e^{2z(x)}  ,
\end{equation}
where we have defined $ \partial_x z(x) = m(x)$,
with the random potential $m(x)$ assumed as before
to be Gaussian with $[m(x) m(x^\prime)]_{\rm ens.} = 2g \delta(x-x^\prime)$.
It remains to specify the boundary conditions on $\psi(x)$
at $x = \pm L/2$.  For technical reasons it is convenient
to ``pin" the 1d random walker at the ends, taking
$z(x=\pm L/2) =0$.

We focus on the ensemble averaged correlation function,
\begin{equation}
  {\cal W}_q(x,L) = [|\psi(x) \psi(0)|^q]_{\rm ens.}  ,
\label{twopoint}
\end{equation}
between two points separated by a distance $x$, assumed much smaller
than $L$.  The one-point function ${\cal W}_q(L) =  {\cal W}_q(x=0,L)$,
referred to as a participation ratio,
is expected to vary as a power law with system size:
\begin{equation}
{\cal W}_q(L) \sim  { {1} \over {L^{d + \tau(q)}}} ,
\end{equation}
with $d=1$ the spatial dimensionality.  The exponent $\tau(q)$ is
often written as $\tau(q) = (q-1)D(q)$.  For a plane wave or
non-fractal wavefunction (as in a 3d metal, say) $D(q) = d$.  For an
exponentially localized wavefunction, $D(q) = \tau(q)=0$.  A simple
fractal would be characterized by a $q$-independent $D$, different
that the spatial dimension, whereas in a multi-fractal $D$ depends on
$q$ and equivalently $\tau(q)$ is a non-linear function of $q$.

The two-point function ${\cal W}_q(x,L)$ for $x$ much larger than microscopic
lengths (ie. lattice spacing) yet much smaller than $L$
is also expected to exhibit power law scaling:
\begin{equation}
{\cal W}_q(x,L) \sim { {1} \over {L^{d + \tau(q)}}} { {1} \over {x^{y(q)}}} ,
\label{exponents}
\end{equation}
with a spectrum of exponents $y(q)$.  For some multi-fractals a
relation can be obtained between $\tau(q)$ and $y(q)$, but in general
they can be independent exponents.

To extract $\tau(q)$ and $y(q)$ for the 1d critical wavefunction, we
follow closely Shelton and Tsvelik.\cite{Shelton97}\ The correlation
function can be expressed as a functional integral over the random
walk configurations $z(x)$ as,
\begin{equation}
  {\cal W}_q(x,L) = {1 \over {Z_0}} \int^\prime \! {\cal D}z \, |\psi(x)
  \psi(0)|^q e^{-S_0}  , 
\end{equation}
with 
\begin{equation}
  Z_0 = \int^\prime\! {\cal D}z  \, e^{-S_0}  ,
\end{equation}
and an action,
\begin{equation}
  S_0 = {1 \over {4g}} \int_{-L/2}^{L/2} \! dx\,  (\partial_x z)^2   .
\end{equation}
Here the prime on the integration indicates the boundary conditions,
$z(x=\pm L/2)=0$.  In the following we put $g=1$.  The functional
integral over disorder configurations is non-trivial due to the
normalization of the wavefunctions.  Fortunately, the normalization
can be exponentiated via the identity,
\begin{equation}
  {\cal N}^{-q} = {1 \over \Gamma(q)} \int_0^\infty \! d\omega\,
  \omega^{q-1} e^{-\omega {\cal N}}   , 
\end{equation}
where it can be absorbed into the action.
In this way one obtains,
\begin{equation}
  {\cal W}_q(x,L) = {1 \over \Gamma(q)} \int_0^\infty \! d\omega\, \omega^{q-1}
  \langle e^{qz(x)} e^{qz(0)}  \rangle  ,
\end{equation}
where the average is given by
\begin{equation}
  \langle e^{qz(x)} e^{qz(0)}  \rangle =
  {1 \over {Z_0}} \int^\prime\! {\cal D}z \, e^{qz(x)} e^{qz(0)} e^{-S}  ,
\end{equation}
with the total action
\begin{equation}
  S =  \int_{-L/2}^{L/2} \! dx\, [{1 \over {4}}(\partial_x z)^2 +
  \omega e^{2z}]  . 
\end{equation}

If $x$ is viewed as an imaginary time coordinate, this average is seen
to be equivalent to a path integral representation of the quantum
mechanics of a particle with coordinate $z$ moving in an exponential
potential.  Passing to the operator representation of this quantum
mechanics by defining a quantum Hamiltonian,
\begin{equation}
  H = - \partial_z^2 + \omega e^{2z}  ,
\end{equation}
the above average can be written
as a quantum expectation value,
\begin{equation}
  \langle e^{qz(x)} e^{qz(0)}  \rangle =
  {  {   \langle 0| e^{-LH/2} e^{qz(x)} e^{qz} e^{-LH/2} |0 \rangle }
    \over
    {  \langle 0| e^{-LH_0} |0 \rangle }  }  ,
  \label{quavg}
\end{equation}
with $e^{qz(x)} = e^{xH} e^{qz} e^{-xH}$, and $H_0 = -\partial_z^2$.
Here $|0 \rangle$ is a position ket with $z=0$.  

Evaluating the wavefunction correlator has thus been reduced to
solving for the quantum mechanics of a particle moving in an
exponential potential -- Liouville quantum mechanics.  This form is
identical to that which arose in the bosonic sector of the
supersymmetric calculation of the Fermion Green's function.  In that
case, the coordinate $z$ was related to the boson number via $n =
e^z$, and $\omega$ was a small imaginary part of the energy.  The
supersymmetric calculation can thus be viewed as a supersymmetric
version of Liouville quantum mechanics.  In earlier work, Kogan, Mudry
and Tsvelik showed that the wave function correlators for a
two-dimensional particle decribed by a Dirac equation with random
vector potential (for which the exact zero energy wavefunction can
also be written down explicitly) could be formulated in terms of
Liouville {\it field theory}.  Perhaps such a 2d localization critical
point can be formulated in terms of a supersymmetric Liouville field
theory.

For simplicity we evaluate the above quantum expectation value within
the hard-wall approximation, which should give the correct scaling
behavior for the wavefunction correlator.  As before, we replace the
exponential potential by a hard-wall at $z_w$, with $\omega e^{2z_w}
=c$, for a constant $c$ of order one.  The value of $c$ affects the
overall prefactor in the correlator.  We choose $c=1/2$ which gives
the correct normalization, ${\cal W}_{q=1}(L) = 1/L$.  Since the
quantum particle is contrained to have $z<z_w$, when $\omega >1$ (and
$z_w$ is negative) the particle's wavefunction vanishes at $z=0$, so
that the quantum expectation value in Eq.~(\ref{quavg}) vanishes
identically.  We can thus restrict the integration over $\omega$ to
$\omega <1$.

The denominator in Eq.~(\ref{quavg}) is the propagator for a free
random walker (i.e without the hard-wall) and can be readily evaluated
giving $(4\pi L)^{-1/2}$.  To evaluate the numerator it is convenient
to let $z \rightarrow z_w - z$, so that the quantum particle is then
constrained to have $z>0$.  Within the hard-wall approximation, the
correlator can then be expressed in terms of the free Hamiltonian as
\begin{eqnarray}
  & & {\cal W}_q(x,L) = { {\sqrt{4\pi L}}  \over {\Gamma(q)2^q} }
  \nonumber \\
  & & \,\, \times \int_0^1 
  { {d\omega} \over \omega} \langle z_w | e^{-LH_0/2} e^{-qz(x)}
  e^{-qz} e^{-LH_0/2} | z_w \rangle  ,
  \label{quexp}
\end{eqnarray}
with $z$ restricted positive and $z_w = |\ln\omega|/2$ (the factor of
two difference between this definition and the one used in the SUSY
calculations is a consequence of a different choice of normalization
of the field $z(x)$).  To evaluate this quantum average we introduce a
complete set of standing waves
\begin{equation}
  \langle z | k \rangle = \sqrt{{2 \over \pi}} \sin(kz),\qquad  (k>0)
\end{equation}
which are eigenstates
$H_0|k \rangle = k^2 |k \rangle$, and appropriately normalized
on the interval $z>0$:  $\langle k|k^\prime \rangle = \delta(k-k^\prime)$.
Inserting the resolution of the identity,
\begin{equation}
1 = \int_0^\infty \! dk\, |k \rangle \langle k|  ,
\end{equation}
into Eq.~(\ref{quexp}), evaluating the matrix elements in closed form
and performing the integration over $\omega$
gives for large $L$ the final result:
\begin{equation}
  {\cal W}_q(x,L) = { {\tilde{\cal W}(q^2 x) } \over { 2^{q-1} q^3
      \Gamma(q) } } { 1 \over L}  ,
\end{equation}
with 
\begin{equation}
  \tilde{\cal W} (x) = { 16 \over \pi}  \int_0^\infty { dk k^2 \over
    (1+k^2)^4  } e^{-xk^2}   .
\end{equation}

The crossover function $\tilde{\cal W} (x)$ interpolates between one
at $x=0$ and $\tilde{\cal W} (x) \sim x^{-3/2}$ for $x >> 1$.  The
crossover length, $x_c = 1/q^2$, is the characteristic distance over
which the logarithm of the wave function $\psi(x)$ changes by order
$1/q$ (or equivalently the ``time" it takes the 1d random walk to move
a distance $\delta z \sim 1/q$).  This crossover scale clearly depends
on the strength of the disorder.  For the original lattice tight
binding model, when the random hopping strengths are comparable to the
mean hopping strength, $\delta t_n /t$ of order one, this crossover
length is of the order of the tight binding lattice spacing.  Thus it is
clear that the {\it form} of the crossover function $\tilde{\cal W}
(x)$ for $x$ of order one, cannot be universal.  In fact, the precise
form above is particular to the hard wall approximation, and an
evaluation using the exact Liouville eignenfunctions would give
another form, although they agree in their universal large $x$
behavior.

For large $x$, ${\cal W}_q(x,L) \sim x^{-3/2} L^{-1}$.  By comparing
with Eq.~(\ref{exponents}), the exponents $\tau(q)$ and $y(q)$ are
seen to be {\it independent} of $q$, with $\tau(q)=0$ and $y(q) =3/2$.
A vanishing $\tau(q)$ is characteristic of an exponentially localized
wavefunction.  However, the two-point correlator for an exponentially
localized wavefunction also decays exponentially ($y(q) = \infty$), in
contrast to the present 1d wavefunction which exhibits power law
correlations.  The 1d critical wavefunction is {\it typically}
quasi-localized (centered around a maximum) with stretched exponential
decay.  However, the {\it average} two-point correlator at separation
$x$ is dominated by the rare wavefunction which has a secondary
maximum close in magnitude but separated spatially (by distance $x$)
from its primary maximum.  The likelihood of this involves the
extremal statistics of a 1d random walk near a global maximum
(absorbing wall), which is being described mathematically above by
quantum mechanics near an exponential (or hard wall) potential.  As
expected, the two-point wavefunction correlator at $q=1$ decays with
the same exponent, $y(q=1)=3/2$, as the average Green's function
obtained with supersymmetry.

\section{Summary and Conclusions}

In this paper we have presented a detailed SUSY analysis of the
critical properties of the zero energy delocalization transition in
the 1d random hopping model.  This 1d random critical point has been
of interest for many years, originating with the pioneering paper by
Dyson in 1953 on a related model of a 1d harmonic chain with random
spring constants.\cite{Dyson}\ Most of the prior work, using a variety
of different approaches, has focussed on properties derivable from the
mean local Green's function, specifically the density of states and
the typical localization length, $\tilde{\xi}$.\cite{McKenzie96}\ By
employing a novel real-space RG approach to analyze phase transtions
in a class of closely related random spin-chains, D.S. Fisher has
recently extended the analysis to extract the spatial dependence of
mean correlation functions.\cite{DFisher94,Fisher95}\ An important
element is the emergence of a {\it second} correlation length, $\xi$,
which determines the spatial decay of {\it mean} (rather than typical)
correlation functions.  To the growing body of knowledge concerning
this 1d random critical point, we add several new results in this
paper.  (i) Using SUSY we have computed the {\it exact} two-parameter
scaling functions for the mean Fermion Green's function.  (ii) By
employing Liouville quantum mechanics, we have extracted the set of
multifractal scaling exponents $\tau(q)$ and $y(q)$ which characterize
the critical wavefunction pair correlators.

Together, these two results encapsulate the important universal
scaling characteristics of this 1d random critical point.  The spatial
dependence of the mean Fermion Green's function is controlled by two
lengths, a mean localization length which diverges upon approaching
the band center as $\xi_{\epsilon} \sim |\ln\epsilon|^2$, and a mean
``staggering length", varying as $\xi_M \sim M^{-2}$ when the
strength, $M$, of a staggering in the hopping strengths is taken to
zero.  These two lengths are to be contrasted with their counterparts,
denoted $\tilde{\xi}_\epsilon$ and $\tilde{\xi}_M$, which charaterize
the spatial decay of the {\it typical} (rather than ensemble-averaged)
Greens function.  From the singular behavior of the density of states,
one can infer that the typical localization length diverges more
slowly, as $\tilde{\xi}_\epsilon \sim |ln \epsilon|$.  Likewise, the
typical staggering length which follows rather directly from the
nature of the exact (decaying) zero energy wavefunction, diverges more
slowly, $\tilde{\xi}_M \sim M^{-1}$, than it's mean counterpart.

For spatial separations $x \ll \xi_\epsilon, \xi_M$, between two points
on the same sublattice, the mean Fermion Green's function varies as an
inverse power law of $x$ with universal exponent $3/2$.  This result
also follows from an analysis of the zero energy wavefunction whose
logarithm undergoes a 1d random walk, with the exponent $3/2$ being
related to {\it extremal} properties of the random walker.  In
contrast, the {\it typical} Green's function for $x \ll
\tilde{\xi}_\epsilon, \tilde{\xi}_M$ is expected to decay as a
stretched exponential, ${\cal G}_{typ} \sim e^{-c\sqrt{x}}$,
reflecting the {\it typical} behavior of the random walker.

A key motivation for the present paper was to investigate in detail
the novel features which emerge in a SUSY formulation of a random
critical point.  The calculation proceeded by expressing mean
correlators in terms of quantum mechanical expectation values for a
SUSY Hamiltonian, which involved a {\it single} superspin.  This
Hamiltonian has a number of notable features: (i) It is non-Hermitian,
with distinct left and right eigenfunctions.  (ii) It has a unique
zero energy ground state, as dictated by supersymmetry, and the
excited states are organized into supersymmetric doublets and
quadruplets.  (iii) The right (or left) eigenstates alone do {\it not}
span the Hilbert space - the Hamiltonian is thus ``defective".  (iv)
The Hilbert space is infinite, due to the non-compact SU(1,1) bosonic
subalgebra of the superspin group.  (v) The eigenstates explore the
outer reaches of the non-compact manifold, in a manner which can be
described by Liouville quantum mechanics.

It is our hope that a thorough undertanding of these unusual features
will be helpful in extending the SUSY approach to attack
two-dimensional random critical points, such as the IQHE plateau
transition.  It is tantalizing to speculate that some appropriate
supersymmetric version of Liouville field theory might give a correct
description of delocalization transitions in 2d.

\acknowledgements We are grateful to Martin Zirnbauer, Claudio Chamon,
Nick Read, and especially Daniel Fisher for useful and enlightening
conversations.  This work has been supported by the National Science
Foundation under grants No. PHY94-07194, DMR-9400142 and DMR-9528578.

\section*{Appendix A}

The mapping from the random transverse field Ising chain to a free 
Fermion model was introduced by Shankar and Murthy.  We briefly 
recapitulate this mapping.  In terms of Majorana Fermions,
\begin{eqnarray}
\eta_{1,n} & = & {1 \over \sqrt{2}} \prod_{m<n} \sigma_m^x \sigma_n^y, 
\\
\eta_{2,n} & = & {1 \over \sqrt{2}} \prod_{m<n} \sigma_m^x \sigma_n^z.
\end{eqnarray} 
which satisfy $\{ \eta_{i,m},\eta_{j,n}\} =
\delta_{ij} \delta_{mn}$,
the random Ising Hamiltonian, Eq.~(\ref{I:hamiltonian})
can be re-written as,
\begin{equation}
{\cal H}_{\rm I} = \sum_n \bigg[ - 2 i K_{1,n} \eta_{1,n} \eta_{2,n} + 2iK_{2,n}
\eta_{1,n} \eta_{2,n+1} \bigg].
\end{equation}
A continuum limit can be taken by putting $x = n dx$,
$K_1 = dx/2$ and $K_2 = (1/2 + m(x))dx$, and converting the sums to
integrals.  This gives
\begin{equation}
{\cal H}_c =  \int dx  \eta \bigg[ \sigma^x i\partial_x + m(x) \sigma^y \bigg] \eta,
\end{equation}
where we have defined a two-component Majorana
field, $\eta = (\eta_1, \eta_2)$.
For spatially uniform $m$ this model describes criticality
in the pure 2d Ising model, with the phase transition occuring at $m=0$.

To complete the mapping it is convenient to
consider a path integral representation of the
partition function, $Z = {\rm Tr} exp(-\beta H)$,
which can be written as a functional integral
over Grassmann fields, $\eta(x,\tau)$,
with associated Euclidian action:
\begin{equation}
S = \int \! dx {{d\omega} \over {2\pi}} \eta(x,\omega) \bigg[ i\omega
+ \sigma^x i\partial_x + m \sigma^y \bigg] \eta(x,-\omega).
\end{equation}
These can be decomposed into new Grassmann fields by defining,
\begin{equation}
\eta_\alpha(\omega) = \overline{\psi}_\alpha (\omega)  ,\quad  
\eta_\alpha(-\omega) = \psi_\alpha (\omega) ,
\end{equation}
for {\it positive} $\omega$ and $\alpha = 1,2$.
In terms of these new fields the action becomes,
\begin{equation}
S = \int_0^\infty {{d \omega} \over {2\pi}} S_\omega ,
\end{equation}
with
\begin{equation}
S_\omega  = \int dx \overline{\psi} \bigg[ \sigma^x i\partial_x + m(x) \sigma^y + i \omega \bigg] \psi .
\end{equation}
Notice that the functional integral factorizes into a product over
independent frequencies.  In the following we focus on only
a single frequency.  The action at a single frequency can be cast
into the form of Eq.~(\ref{Hamiltonian}) 
by a rotation in ``spin-space" around the y-axis by $\pi/2$,
which takes $\sigma^x \rightarrow \sigma^z$,
giving
\begin{equation}
S_\omega = \int dx \overline{\psi} (h + i\omega) \psi  ,
\end{equation}
with $h$ the 1d random Hamiltonian in Eq.~(\ref{hamiltonian}).
Note that, in this case, a non-zero mass $m_0$ corresponds simply to the 
deviation from the Ising critical point.

\section*{Appendix B}

For the special case of the ground state
wavefunction, we are in fact able to obtain an exact solution {\sl
  without taking the continuum limit}.  This is possible because the
difference equation, Eq.~(\ref{SchrodE0}), is linear in $n$.  Here, we
specialize to the case $n=0$, in which this solution is especially simple.

For $M=0$, the Schr\"odinger equation decouples on even and odd
sublattices and can be solved independently on each.  To bring this
out, we define
\begin{equation}
  \phi_{2n} = \gamma_n^e, \qquad \phi_{2n+1} = \gamma_n^o,
\end{equation}
for $n=0,1,2,\ldots$.  The even and odd sublattice fields then obey
\begin{eqnarray}
  \omega\gamma^o_n = (n+1)\gamma^o_{n+1} -
  (2n+1)\gamma^o_n + n\gamma^o_{n-1}  & \qquad & n \geq 0,
  \nonumber \\ 
  \omega\gamma^e_n  = (n+1/2)\gamma^e_{n+1} -
  2n\gamma^e_n + (n-1/2)\gamma^e_{n-1} & \qquad & n > 0. \nonumber
\end{eqnarray}

To solve them, we define the generating function
\begin{equation}
  \hat\gamma^P(w) = \sum_{n=0}^\infty \gamma^P_n w^n,
\end{equation}
where $P=o,e$.  Consider first the odd sector.  Multiplying the
equation for $\gamma^0_n$ by $w^n$
and summing gives 
\begin{equation}
  (1-w)^2 {{d\hat\gamma^o} \over {dw}} = (1-w+\omega)\hat\gamma^o
\end{equation}
This is easily solved by separation of variables, to give
\begin{equation}
\hat\gamma^o = {C \over {1-w}}\exp\left[ {\omega \over {1-w}} 
\right],
\end{equation}
where $C$ is an arbitrary constant.  Not the strong divergence at $w=1$.  
This implies unacceptable behavior for $\gamma^o_n$ at large $n$.  

The even sector is (fortunately!) rather more complicated.  The
crucial difference is the fact that Eq.~(\ref{SchrodE0}) is valid only
for $n>0$, leaving an extra free parameter.  Carrying out the
transform in this case gives
\begin{eqnarray}
  (1-w)^2 {{d\hat\gamma^e} \over {dw}} +&& \left[{1 \over 2}\left(w
       - {1 \over w}\right) - \omega\right]\hat\gamma^e \nonumber \\
  & & \;\;\; = -\left[{1 \over 
      2}\left({1\over w} - \gamma^e_1\right) + \omega\right],
\end{eqnarray}
Where we have imposed the normalization $\gamma^e_0 = 1$.  Note the
appearance of $\gamma^e_1$ as a parameter in the equation.  It must be
adjusted to achieve a well-behaved solution.

This inhomogenous equation can be solved by introducing the integrating 
factor $(1-w)/\sqrt{w} \exp[-\omega/(1-w)]$.  The solution is
\begin{eqnarray}
  {{1-w} \over \sqrt{w}}&& e^{-\omega/(1-w)} \hat\gamma^e(w) =  
  \nonumber \\
  & & - \int^w {{dy} \over{\sqrt{y}(1-y)}} \left[ {{1-y} \over {2y}} + 
    {\delta \over 2} + \omega\right] e^{-\omega/(1-y)},
\end{eqnarray}
where $\delta = 1 - \gamma^e_1$.  Performing an integration by 
parts leads to the form
\begin{equation}
  \hat\gamma^e(w) = {1 \over {1-w}}\left[1+\sqrt{w}e^{\omega/(1-w)} 
    J(w)\right],
\end{equation}
where
\begin{equation}
  J(w) = \int_0^w {{dy} \over{\sqrt{y}(1-y)}} \left[{\omega\over {1-y}} - 
    {\delta \over 2} - \omega\right] e^{-\omega/(1-y)}.
\end{equation}
To avoid the strong divergence of $\gamma^e_n$, we clearly need 
\begin{equation}
  J(1) = 0.
\end{equation}
This fixes $\delta$.  This implies that
\begin{equation}
  J(w) = - \int_w^1 {{dy} \over{\sqrt{y}(1-y)}} \left[{\omega\over {1-y}} - 
    {\delta \over 2} - \omega\right] e^{-\omega/(1-y)}.
\end{equation}
We now change variables via $y = 1-\omega/t$, and also define 
$s=(1-w)/\omega$.  Then
\begin{equation}
  J = -\int_{1/s}^\infty {{dt} \over {t\sqrt{1-\omega/t}}} \left[t - 
    {\delta \over 2} - \omega\right]e^{-t}.
\end{equation}
We may now take the limit $\omega \rightarrow 0$, with $s$ fixed, and 
$\delta \gg \omega$.  Then
\begin{equation}
  J(s) \rightarrow {\delta \over 2}E_1(1/s) - e^{-1/s},
\end{equation}
where
\begin{equation}
  E_1(x) = \int_x^\infty dt e^{-t}/t
\end{equation}
is the exponential-integral function.  Plugging back in gives, finally,
\begin{equation}
  \hat\gamma^e(s) = {\delta \over {2\omega s}}e^{1/s} E_1(1/s) = {\delta \over 
    {2\omega}} \int_0^\infty dt {{e^{-t}} \over {1+ts}}.
\end{equation}
In fact, $\hat\gamma^e(s)$ is nothing but the Laplace transform in 
the limit $\omega \ll 1$,
\begin{eqnarray}
  \hat\gamma^e(s) & = & \sum_n \gamma^e_n w^n = \sum_n 
  \gamma^e_n e^{-n\omega s} \nonumber \\
  & \rightarrow & \int dn \gamma^e (n) e^{-n \omega s} = 
  [L\gamma^e](\omega s).
\end{eqnarray}
We can therefore invert it using the inversion formula
\begin{eqnarray}
  \gamma^e_n & = & \int_{c-i\infty}^{c+i\infty} {{ds} \over {2\pi i}} e^{ns} 
  L\gamma^e(s) \nonumber \\
  & = & \omega \int_{c-i\infty}^{c+i\infty} {{ds} \over {2\pi i}}
  e^{n\omega s}  
  [L\gamma^e](\omega s)  \nonumber \\
  & = & {\delta \over 2} \int_0^\infty dt e^{-t} \int 
  _{c-i\infty}^{c+i\infty} {{ds} \over {2\pi i}} {e^{n\omega s} \over 
    {1+ts}} \nonumber \\
  & = & {\delta \over 2} \int_0^\infty dt {1 \over t} e^{-n\omega/t -t} 
  \nonumber \\
  & = & {\delta \over 2} \int_0^\infty {{dt} \over t} e^{-n t -\omega/t}.
\end{eqnarray}
This is precisely the modified Bessel function solution obtained 
from the continuum limit (see appendix E).

\section*{Appendix C}

In this Appendix we obtain the density of states
without resorting to the hard-wall approximation, by solving
exactly the full continuum equation Eq.~(\ref{continuum_eqn}).
This can be accomplished by employing an inverse Laplace transform, defining
\begin{equation}
  \phi(n,M) = \int_0^\infty dt e^{-nt} \tilde\phi(t). \label{ILT}
\end{equation}
Provided $t^2\tilde\phi(t) \rightarrow 0$ as $t\rightarrow 0$ (as
required for a well-behaved solution as $n\rightarrow\infty$),
insertion into Eq.~(\ref{continuum_eqn}) leads to the simple transformed
form
\begin{equation}
  t^2 {{d\tilde\phi} \over {dt}} + (1 - M) t \tilde\phi =
  {\omega \over 2}\tilde\phi . 
\end{equation}
This has the general solution
\begin{equation}
  \tilde\phi(t) = {a \over t^{1-M}} e^{-\omega/2t},
\end{equation}
where $a$ is an arbitrary constant.
The ``unnormalized" wavefunction is thus given by
\begin{equation}
  \phi(n,M) = a\int_0^\infty {{dt} \over {t^{1-M}}} 
  e^{-nt-\omega/2t} ,
  \label{intrep}
\end{equation}
which is the integral representation of a Bessel function, $\phi(n,M)
= 2a (\omega/2n)^{M/2}K_{M}(\sqrt{2\omega n})$.  To evaluate
Eqs.~(\ref{constraint1},\ref{constraint2}), we need $\phi(1)$ and
$\phi'(1)$.  These are determined by making the change of variables
$x=t^{-M}$, which yields
\begin{equation}
\phi(1,M) =  {a \over M} \int_0^\infty \!\!\! dx \exp\left[ - 
    x^{1/{M}}\! -\! \left( x W^{-1}\right)^{-1/{M}} \right] ,
\label{nice_integral_form}
\end{equation}
with the scaling variable $W = (\omega/2)^M$.  In the scaling limit
$\omega,{M} \rightarrow 0$ with $W$ fixed and finite,
Eq.~(\ref{nice_integral_form}) can be simply evaluated.  Since each of
the arguments in the exponential goes to zero or infinity, the limits
of integration are restricted giving,
\begin{equation}
  \phi(n=1,M) = {a \over  M} (1- \omega^{M}),
\end{equation}
(using $2^M \approx 1$).  The same change of variables can be used to
extract the $n$ derivative, giving in the scaling limit
\begin{equation}
{{d \phi} \over {d n}} |_{n=1} = - a .
\end{equation}
Comparison with the hard-wall forms shows that the constants $c_1$ and
$c_2$ are identical provided we take $a= 1/z_w = |\ln\omega|^{-1}$.

With the exact solutions of the continuum equations in hand, one can
readily evaluate the density of states by inserting the integral
representations Eq.~(\ref{intrep}), into the expression for
$G(i\omega)$ in Eq.~(\ref{intDOS}).  The $n$-integration can be
readily performed.  In the scaling limit $M \ll 1$ the remaining two
$t$-integrations are simple and yield an identical result to
Eqs.~(\ref{exactG}--\ref{hard_wall_sf}).

\section*{Appendix D}

In this appendix, we obtain the exact excited state wavefunctions in
the continuum limt in the $N_F=0,2$, $N_B = -1$ sectors, and show that
they lead to the same scaling form for the Green's function as does
the hard-wall approximation.  Beginning with
Eq.~(\ref{continuum_singlet}), we make the change of variables
\begin{equation}
  \chi(n,M) = a(\beta)n^{-(1+M)/2} \tilde{\chi}(n,\beta),
\end{equation}
where $a(\beta)$ is a normalization constant to be chosen later in
order to maintain the closest possible agreement with the hard-wall
solutions in section V.
The transformed wavefunction then satisfies the simpler equation
\begin{equation}
  \bigg[ n^2 {d^2 \over {dn^2}} + n {d \over {dn}} + {1 \over
    4}\left(\beta^2 - 2\omega n\right)\bigg] \tilde\chi(n),
  \label{chi_0_eqn}
\end{equation}
where $\beta = \sqrt{E-M^2}$.   Eq.~(\ref{chi_0_eqn}) is a standard
equation of classical mathematical physics.  Its solutions are
modified Bessel functions of imaginary index:
\begin{equation}
  \tilde\chi(n) = K_{i\beta}(\sqrt{2n\omega}),
\end{equation}
where we have chosen the solution $K$ which decays at infinity.  Note
that we have assumed $E \geq M^2$, for which $\beta$ is real.  It is
straightforward to show that there are no satisfactory solutions with
$E < M^2$.  Very few results are readily available for these functions
at imaginary index.  Instead, we will make heavy use of the integral
representation,
\begin{equation}
  \tilde\chi(n,\beta) = \int_0^\infty e^{-\sqrt{2n\omega} \cosh t} \cos
  \beta t,
  \label{chi_0t_sol}
\end{equation}
which can be verified by direct substitution into
Eq.~(\ref{chi_0_eqn}).  A second useful form is obtained by
integrating Eq.~(\ref{chi_0t_sol}) by parts:
\begin{equation}
  \tilde\chi(n,\beta) = {\sqrt{2n\omega} \over \beta} \int_0^\infty
  e^{-\sqrt{2n\omega} \cosh t} \sinh t \sin \beta t.
  \label{second_form}
\end{equation}
The first task at hand is to determine the spectrum, or allowed values
of $\beta$.  To do this required asymptotic matching for the continuum
solution (valid for $n \gg 1$) to the ``outer'' solution of the
difference equation with $\omega=0$ (valid for $n \ll 1/\omega$).
This matching is imposed in the overlap region $1 \ll n \ll
1/\omega$.  To study this limit, we let $s = e^t$ in
Eq.~(\ref{second_form}), which is then dominated by $s \gg 1$.  Thus
\begin{equation}
  \tilde\chi(n,\beta) \sim \sqrt{n\omega/2} {1 \over {2i\beta}}
  \int_1^\infty ds \left(s^{i\beta}-s^{-i\beta}\right)
  e^{-\sqrt{n\omega/2} s}.
\end{equation}
This gives, 
\begin{equation}
  \chi(n,M) \sim {1 \over {2i\beta}} n^{-(1+M)/2}
  \left[(n\omega)^{-i\beta/2} - (n\omega)^{i\beta/2}\right].
\end{equation}
Comparison to the outer solution, Eq.~(\ref{outer}), 
\begin{equation}
  \chi(n,M) \sim n^{-(1+M)/2} \left[ c_+ n^{-i\beta/2} + c_-
    n^{i\beta/2}\right], 
\end{equation}
then gives, as in section V,
\begin{eqnarray}
  \beta_k & = & \pi k/ z_w, \\
  c_3 & = & (-1)^{k+1} c_4 \equiv c.
\end{eqnarray}

We must next determine the constants $c$ and $a(\beta)$.
Normalization requires
\begin{equation}
  2|C_k|^2 |a(\beta)|^2 I_\beta = 1,
\end{equation}
where
\begin{equation}
  I_\beta = \int_1^\infty {{dn} \over n} |\tilde\chi(n,\beta)|^2.
\end{equation}
\end{multicols}
Performing the integral over $n$ gives
\begin{equation}
   I_\beta = {2 \over \beta^2} \int_0^\infty \!\! dt dt'
   \sin \beta t 
  \sin \beta t' {{\sinh t \sinh t'} \over {(\cosh t + \cosh t')^2}}
  \bigg[1 + \sqrt{2\omega}(\cosh t + \cosh
      t')\bigg] e^{-\sqrt{2\omega}(\cosh t + \cosh t')}.
\end{equation}
The next step is to rescale the parameter $t \rightarrow t/\beta$, $t'
\rightarrow r'/\beta$ to give
\begin{equation}
  I_\beta = {1 \over {2\beta^4}} \int_0^\infty \!\! dt dt' \sin t
  \sin t' {\rm sech}^2 \left({{t-t'} \over {2\beta}}\right)
   \bigg[1 +\sqrt{2\omega}(\cosh t/\beta + \cosh
    t'/\beta)\bigg] e^{-\sqrt{2\omega}(\cosh t/\beta + \cosh t'/\beta)}.
  \label{rescaled_norm}
\end{equation}
\begin{multicols}{2}
We are interested in small $\omega$, with $\beta = \pi
k/|\ln(\omega)|$.  In this limit,
\begin{equation}
  \cosh t/\beta = {1 \over 2}\left( e^{t/\beta} + e^{-t/\beta}\right)
  \approx {1 \over 2} \omega^{-{t \over {\pi k}}},
\end{equation}
and an identical result with $t \rightarrow t'$.  The factor in the
exponential in Eq.~(\ref{rescaled_norm}) therefore becomes
\begin{equation}
  \sqrt{2\omega}(\cosh t + \cosh t') \stackrel{\longrightarrow}{\omega 
  \rightarrow 0}  \cases{0 & $ 0 < t,t' < \pi k/2$ \cr \infty &
  otherwise \cr}.
\end{equation}
Taking this limit therefore acts simply to restrict the limits of
integration, and we have
\begin{equation}
  I_\beta = {1 \over {2\beta^4}} \int_0^{\pi k/2} \!\! dt dt' \sin t \sin
  t' {\rm sech}^2  \left({{t-t'} \over {2\beta}}\right).
\end{equation}
Since the ${\rm sech}$ is sharply peaked around zero in the $\beta
\rightarrow 0$ limit, we may effectively set $t' \approx t$ in the
second sine to obtain
\begin{equation}
  I_\beta= {1 \over {2\beta^4}} \int_0^{\pi k/2} dt \sin^2 t
  \int_0^{\pi k/2} \! dt' {\rm sech}^2 \left({{t-t'} \over {2\beta}}\right).
\end{equation}
The $t'$ integral is clearly proportional to $\beta$, and performing
these integrations exactly gives, in the $\beta \rightarrow 0$ limit,
\begin{equation}
  I_\beta = {{\pi k} \over {2\beta^3}}.
\end{equation}
Thus, by choosing
\begin{equation}
  a(\beta) = \beta = {{\pi k} \over {|\ln\omega|}},
\end{equation} 
we obtain the same constant $c = z_w^{-1/2}$ as found for the
hard-wall solutions in section V.

We are now in a position to calculate the Fermion Green's function.  
As in section V, to use the decomposition in 
Eq.~(\ref{GF_expansion}), we must calculate matrix elements of 
single-Fermion operators between the ground and excited states.  In 
general, using Eqs.~(\ref{superposition},\ref{chi_decomposition}), 
\end{multicols}
\begin{eqnarray}
        \left._{L}\langle\right. k | F_{\beta}^{\dagger} |0 \rangle_{R} 
        & \sim & {{c_{1}} \over \sqrt{2z_w}} \int_{1}^{\infty}
        {{dn} \over  
        {\sqrt{n}}} \Bigg[ (-1)^{\beta} \phi(n,M) \chi(n,-M) + (-1)^{k+1} 
        \phi(n,-M) \chi(n,M) \Bigg] \\
        & = & {{c_{1}} \over \sqrt{2z_w}} \Bigg[ (-1)^{\beta} I(M) + 
        (-1)^{k+1} I(-M) \Bigg],
        \label{mat_elt}
\end{eqnarray}
\begin{multicols}{2}
where the integral
\begin{equation}
        I(M) = {{\pi k} \over z_w}\int_1^{\infty} {{dn} \over
          {\sqrt{n}}} \phi(n,M)  
        \tilde\chi(n,\beta) n^{M/2}.
\end{equation}
To evaluate this integral, we let $t\rightarrow e^{t}$ in the integral 
representation, Eq.~(\ref{intrep}), giving
\begin{equation}
        \phi(n,M) = c_{1} {\omega \over n}^{M/2} \int_{-\infty}^{\infty} \! 
        dt \; e^{-\sqrt{2n\omega} \cosh t} e^{Mt},
        \label{phirep2}
\end{equation}
Inserting this and Eq.~(\ref{chi_0t_sol}) above, the $n$--integration 
can be readily performed, giving
\begin{equation}
        I(M) = {{\pi k\omega^{M/2}} \over {\sqrt{2\omega} z_w^2}} 
        \int_{-\infty}^{\infty} \! dt_{1} dt_{2} {{e^{Mt_1} \cos \beta t_2} 
        \over {\cosh t_1 + \cosh t_2}}.
\end{equation}
The limits $M,\beta \rightarrow 0$ can be safely taken in the 
numerator of the integral.  The final result is
\begin{equation}
        I(M) = {{\pi^3 k \omega^{-(1-M)/2}} \over {\sqrt{2} z_w^2}}.
\end{equation}
Putting this into Eq.~(\ref{mat_elt}) above and thence into 
Eq.~(\ref{GF_expansion}), one recovers the final result, 
Eqs.~(\ref{G_omega_scaling}--\ref{length_scales}) quoted in 
section V, with a different value, $A=\pi^6/4$, for the nonuniversal 
constant.

\section*{Appendix E}

In this appendix, we solve the difference equation for ${\cal J}^x$
eigenstates, Eq.~(\ref{Jx_eigenstates}), in the appropriate sector for
the fermion Green's function.  Consider the generating function,
\begin{equation}
  \hat\psi(w,\alpha) = \sum_{n=0}^\infty \psi_n(\alpha) w^n.
\end{equation}
Multiplying Eq.~(\ref{Jx_eigenstates}) by $w^n$ and summing gives
\begin{equation}
  (1-w^2){d \over {dw}}\hat\psi(w,\alpha)  - w\hat\psi(w,\alpha) =
  -2i\alpha\hat\psi(w,\alpha).
\end{equation}
This is easily solved by separation of variables.  One finds
\begin{equation}
  \hat\psi(w,\alpha) = (1+w)^{-1/2-i\alpha} (1-w)^{-1/2+i\alpha},
\end{equation}
choosing $\psi_0(\alpha)=1$ to fix the overall constant.  This can be
inverted using the contour integral
\begin{equation}
  \psi_n(\alpha) = \oint {{dw} \over {2\pi i}} {{\hat\psi(w,\alpha)}
    \over {w^{1+n}}}.
\end{equation}
\end{multicols}
Deforming the contour to obtain a real integral gives
\begin{equation}
  \psi_n(\alpha) = {{\cosh \pi\alpha} \over \pi} \int_1^\infty {{dw}
    \over {w^{1+n}}} \left[ (w+1)^{-1/2-i\alpha}(w-1)^{-1/2+i\alpha} +
    (-1)^n (w+1)^{-1/2+i\alpha} (w-1)^{-1/2-i\alpha} \right].
\end{equation}
\begin{multicols}{2}
  For large $n \gg 1$, this integral is dominated by $w \approx 1$,
  and can be easily evaluated to give the result quoted in
  Eq.~(\ref{Jx_asymptotic}).

\end{multicols}

\begin{references}

\bibitem{Kramer93}
B. Kramer and A. MacKinnon, Rep. Prog. Phys. {\bf 56},  1469  (1993),
and references therein.

\bibitem{Huckestein95}
B. Huckestein, Rev. Mod. Phys. {\bf 67},  357  (1995), and references therein.


\bibitem{Schaefer80}
L. Sch\"aefer and F. Wegner, Z. Phys. B {\bf 38}, 113 (1980).

\bibitem{Hikami81}
S. Hikami, Phys. Rev. B {\bf 24},  2671  (1981).

\bibitem{Zirnbauer86}
M.~R. Zirnbauer, Nuc. Phys. B {\bf 265},  375  (1986).

\bibitem{Chalker88}
J.~T. Chalker and P.~D. Coddington, J. Phys. C {\bf 21},  2665  (1988).

\bibitem{Chalker88a}
J.~T. Chalker and G.~J. Daniell, Phys. Rev. Lett. {\bf 61},  593  (1988).

\bibitem{Efetov83}
K.~B. Efetov, Adv. Phys. {\bf 32},  53  (1983).

\bibitem{Zirnbauer94}
M.~R. Zirnbauer, Annalen der Physik {\bf 3},  513  (1994).

\bibitem{Lee94} A similar mapping using replicas was established in
  D.~H. Lee, Phys. Rev. B {\bf 50}, 10788 (1994).

\bibitem{Kondev96}
J. Kondev and J.~B. Marston, cond-mat/9612223 (unpublished).

\bibitem{Balents97}
L. Balents, M.~P.~A. Fisher, and M.~R. Zirnbauer, Nucl. Phys. B {\bf 483},  681
   (1997);  I.A. Gruzberg, N. Read and S. Sachdev, Phys. Rev. B {\bf
     55}, 10593 (1997); and cond-mat/9704032 (unpublished).

\bibitem{Falko95}
V.~I. Fal'ko and K.~B. Efetov, Phys. Rev. B {\bf 52},  17413  (1995).

\bibitem{Ludwig94}
A.~W.~W. Ludwig, M.~P.~A. Fisher, R. Shankar, and G. Grinstein, Phys. Rev. B
  {\bf 50},  7526  (1994).

\bibitem{Mudry96}
C. Mudry, C. Chamon, and X.-G. Wen, Nucl. Phys. B {\bf 466},  383  (1996).

\bibitem{Chamon96}
C.~d.~C. Chamon, C. Mudry, and X.-G. Wen, Phys. Rev. Lett. {\bf 77},  4194
  (1996).

\bibitem{Kogan96}
I.~I. Kogan, C. Mudry, and A.~M. Tsvelik, Phys. Rev. Lett. {\bf 77},  707
  (1996).

\bibitem{McKenzie96}
For a survey of previous work, see R.~H. McKenzie,
Phys. Rev. Lett. {\bf 77},  4804  (1996), and references therein.

\bibitem{DFisher94}
D.~S. Fisher, Phys. Rev. B {\bf 50},  3799  (1994).

\bibitem{Fisher95}
D.~S. Fisher, Phys. Rev. B {\bf 51},  6411  (1995).

\bibitem{DSFunpub} D. S. Fisher, private communication (unpublished) (1997).

\bibitem{Shelton97}
D.~G. Shelton and A.~M. Tsvelik, cond-mat/9704115 (unpublished).

\bibitem{Dyson}
F. Dyson, Physical Review, {\bf 92}, 1331 (1953).

\bibitem{Thouless72}
D.~J. Thouless, J. Phys. C {\bf 5},  77  (1972).

\end{references}
\end{document}